\documentclass[lettersize,journal]{IEEEtran}
\usepackage{amsmath,amsfonts}
\usepackage{algorithmic}
\usepackage{algorithm}
\usepackage{array}
\usepackage[caption=false,font=normalsize,labelfont=sf,textfont=sf]{subfig}
\usepackage{textcomp}
\usepackage{stfloats}
\usepackage{url}
\usepackage{verbatim}
\usepackage{graphicx}
\usepackage{multirow}
\usepackage[flushleft]{threeparttable}
\begin{document}

\title{Operational Neural Networks for Parameter-Efficient Hyperspectral Single-Image Super-Resolution}
\author{Alexander Ulrichsen,~\IEEEmembership{Student Member,~IEEE,} Paul Murray,~\IEEEmembership{Member,~IEEE,} Stephen Marshall,~\IEEEmembership{Senior Member,~IEEE,} Moncef Gabbouj~\IEEEmembership{Fellow,~IEEE,} Serkan Kiranyaz,~\IEEEmembership{Senior Member,~IEEE,} Mehmet Yamaç, Nour Aburaed,~\IEEEmembership{Graduate Student Member,~IEEE}
\thanks{A. Ulrichsen is with the Department of Electronic and Electrical Engineering, University of Strathclyde, Glasgow, UK}
\thanks{P. Murray is with the Department of Electronic and Electrical Engineering, University of Strathclyde, Glasgow, UK}
\thanks{S. Marshall is with the Department of Electronic and Electrical Engineering, University of Strathclyde, Glasgow, UK}
\thanks{M. Gabbouj is with the Faculty of Information Technology and Communication Sciences, Tampere University, Tampere, Finland}
\thanks{S. Kiranyaz is with the Electrical Engineering Department, College of Engineering, Qatar University, Qatar}
\thanks{M. Yamaç is with the Faculty of Information Technology and Communication Sciences, Tampere University, Tampere, Finland}
\thanks{N. Aburaed is with the Department of Electronic and Electrical Engineering, University of Strathclyde, Glasgow, UK, and the MBRSC Lab, University of Dubai, Dubai, UAE}}

\markboth{IEEE Journal of Selected Topics in Applied Earth Observations and Remote Sensing,~Vol.~XX, No.~X, Month~Year}%
{Shell \MakeLowercase{\textit{et al.}}: A Sample Article Using IEEEtran.cls for IEEE Journals}

\IEEEpubid{0000--0000/00\$00.00~\copyright~2021 IEEE}

\maketitle

\begin{abstract}
Hyperspectral Imaging is a crucial tool in remote sensing which captures far more spectral information than standard color images. However, the increase in spectral information comes at the cost of spatial resolution. Super-resolution is a popular technique where the goal is to generate a high-resolution version of a given low-resolution input. The majority of modern super-resolution approaches use convolutional neural networks. However, convolution itself is a linear operation and the networks rely on the non-linear activation functions after each layer to provide the necessary non-linearity to learn the complex underlying function. This means that convolutional neural networks tend to be very deep to achieve the desired results. Recently, self-organized operational neural networks have been proposed that aim to overcome this limitation by replacing the convolutional filters with learnable non-linear functions through the use of MacLaurin series expansions. This work focuses on extending the convolutional filters of a popular super-resolution model to more powerful operational filters to enhance the model performance on hyperspectral images. We also investigate the effects that residual connections and different normalization types have on this type of enhanced network. Despite having fewer parameters than their convolutional network equivalents, our results show that operational neural networks achieve superior super-resolution performance on small hyperspectral image datasets. Our code is made available on Github: https://github.com/aulrichsen/SRONN.
\end{abstract}

\begin{IEEEkeywords}
Hyperspectral Imaging, Super-Resolution, Operational Neural Networks
\end{IEEEkeywords}

%\titlepgskip=-15pt

\section{Introduction}
\label{sec:introduction}
\IEEEPARstart{H}{yperspectral} imaging is a key tool in remote sensing applications such as material classification, mineral exploration, environmental monitoring, and more \cite{HS-RS}. The reason it is valuable is due to its additional spectral information which offers insights into the materials within the image that standard color images cannot provide. However, due to sensor limitations, it is difficult to obtain a high-quality hyperspectral image (HSI) with both high spectral and spatial resolution \cite{villafranca2012limitations} and thus the increased spectral resolution comes at the cost of decreased spatial resolution \cite{brady2009optical}. Automated image processing tasks such as image segmentation, object detection and classification can improve the efficiency of remote sensing systems. However, the reduction in spatial resolution can be detrimental to their performance. It is therefore desirable to be able to recover the lost spatial resolution to improve the performance of post-processing tasks on the resulting HSI. Single image super-resolution (SISR) is a technique used to enhance the spatial resolution of the given low-resolution hyperspectral image without any auxiliary information. 

Most modern super-resolution (SR) approaches use convolutional neural networks (CNNs) to produce an image-to-image mapping operator which converts the input low-resolution image to a high-resolution image \cite{wang2021real, wang2022semi, dong2015image, singh2022semi}. These operators are of a complex non-linear nature and part of the reason that CNNs have had so much success in this field is due to their capacity to learn complex non-linear operators. However, the sole non-linear elements of a CNN come from the activation functions after each layer, meaning that CNNs often require many layers to have the necessary non-linear capacity and diversity to learn the desired operator. Recently, operational neural networks (ONNs) \cite{kiranyaz2020operational, malik2020fastonn} and their new variants, self-organised operational neural networks (Self-ONNs) \cite{kiranyaz2021self}, have been proposed to overcome this limitation by using the generative neuron model that can customize the optimal non-linear function during training for each kernel element. To accomplish this, each kernel element is extended with MacLaurin series expansions and the terms of the series are made learnable. This means that each kernel element can learn to approximate any non-linear function and thus similar theoretical non-linear capacity of a deep CNN can be achieved in a much shallower Self-ONN which is more computationally efficient.
\IEEEpubidadjcol	% Allow space for pub id in the second column
In this paper, we take the popular SR network, SRCNN \cite{dong2015image}, and extend it for use on hyperspectral images. We also make a Self-ONN equivalent model by replacing the convolutional layers with operational layers. Furthermore, we make a Self-ONN version with a reduced number of filters to demonstrate the non-linear capacity of operational layers over convolutional layers. We train our models on the publicly available Pavia University, Cuprite, Salinas, and Urban datasets \cite{rslab, HS-RSS} and show that Self-ONNs can provide a HSI SR performance improvement of over 0.5 dB PSNR even when it has fewer parameters than a CNN with an equivalent architecture.

Furthermore, this study investigates the effects residual connections and various normalization types have on Self-ONN performance, as, to the best of our knowledge, this has not been previously investigated.

The novel and significant contributions of this study can be summarised as follows:
\begin{itemize}
  \item Based on the SRCNN \cite{dong2015image} configuration, novel Self-ONNs have been proposed for the Hyperspectral Single-Image Super-Resolution task.
  \item We incorporate residual connections and various normalization layers into Self-ONN models, which to the best of our knowledge, has never been done before, and present our novel findings on the performance effects these layer types have on our Self-ONN models.
  \item With the proposed model and structural modifications, we have achieved performance improvements with a reduced number of overall network parameters compared to the SRCNN model.
\end{itemize}

The rest of the paper is organized as follows: Section \ref{sec:relatedWork} will briefly present the related work with the conventional ONNs. Section \ref{sec:methodology} details the proposed methodology for Hyperspectral Single-Image Super-Resolution. We present the experimental setup and results in Section \ref{sec:results} along with detailed comparative evaluations in Section \ref{sec:discussion}. Finally, Section \ref{sec:conclusion} concludes the paper and suggests topics for future research.

\section{Related Work}\label{sec:relatedWork}

%\subsection{Hyperspectral Imaging}
%\noindent Hyperspectral imaging is a valuable tool in remote sensing applications such as material classification, mineral exploration, environmental monitoring, and more \cite{HS-RS}. The effectiveness of these tasks is dependent on the resolution of the captured image. However, due to sensor limitations, it is difficult to obtain a high-quality hyperspectral image (HSI) with both high spectral and spatial resolution \cite{villafranca2012limitations} and thus the increased spectral resolution generally comes at the cost of decreased spatial resolution \cite{brady2009optical}. It is therefore desirable to be able to recover the lost spatial resolution to improve the performance of post-processing tasks.

\subsection{Super-Resolution}
\noindent Most modern approaches to super-resolution use convolutional neural networks (CNNs) in either a supervised or unsupervised manner \cite{lim2017enhanced, shi2016real, ledig2017photo, yamac2021kernelnet}. Supervised training involves training a model on a dataset consisting of low-resolution and high-resolution image pairs. One of the first papers to adopt this approach was \cite{dong2015image} where they proposed their CNN model named SRCNN for the task of single image super-resolution. 

SRCNN is a fairly shallow CNN consisting of only 3 layers, so the authors of \cite{kim2016accurate} proposed a much deeper CNN to perform supervised SISR. The deeper network provides more learning capacity but is also more difficult to train due to the vanishing gradient problem. To overcome this issue, they proposed a residual connection which sums the input of the model directly to the output so that instead of learning the direct input-to-output image mapping, the model learns the residual between the input and output which improved results and greatly decreased training times. 

Since then, many other deep CNN models have been proposed for supervised single image super-resolution \cite{arun2020cnn, zheng2018multi, song2018super, lim2017enhanced, zhao2016loss}. However, the main challenge of this approach is acquiring the dataset. Ideally, perfectly aligned images would be captured with a low-resolution and a high-resolution sensor, but this is impractical to perform in many situations. What is more commonly done is a dataset of high-resolution images is acquired and the low-resolution image pairs are then synthetically generated by blurring and downsampling the high-resolution images and then adding noise.

To overcome this limitation, unsupervised methods using Generative Adversarial Networks (GANs) \cite{goodfellow2014generative} have been proposed which utilize datasets of unpaired real high-resolution and low-resolution images through the use of generator and discriminator models. The generator produces high-resolution versions of the low-resolution images and the discriminator aims to distinguish between the true high-resolution images and the generated high-resolution images. Over time, the generator learns to produce realistic high-resolution outputs of the input low-resolution images which match the distribution of the high-resolution image dataset. Thus, the model is more likely to learn the true low-resolution to high-resolution image mapping function. Many researchers have achieved impressive results using this approach \cite{wang2021real, ignatov2018wespe, kim2020unsupervised, ledig2017photo, huang2019hyperspectral, bell2019blind}. However, the unsupervised nature of this approach means that it is inherently more difficult to train as the generator learns from feedback provided by the discriminator and the discriminator has no prior knowledge of the objective. In addition, it is also challenging to measure the performance of a GAN objectively as typical image quality metrics such as peak signal-to-noise ratio (PSNR) and structural similarity index (SSIM) \cite{wang2004image} require a target image to be evaluated. To overcome these problems it typically requires a lot of training data to produce a realistic GAN \cite{karras2020training} which is not always available, particularly in the case of hyperspectral imagery.

Attempts have been made to improve upon human-perceived super-resolution quality. \cite{johnson2016perceptual} introduces a perceptual loss function generated through a fixed loss network to create visually pleasing results but at the cost of PSNR and SSIM, indicating that their per-pixel accuracy is lower. \cite{ledig2017photo} introduces a perceptual loss function to train a GAN, which again focuses on learning mappings that are perceptually pleasing to humans, rather than pixel-to-pixel accuracy. These approaches improve how pleasing super-resolution outputs may be to a human observer, but do not necessarily provide any performance improvement for post-processing tasks to be done on the resulting images.

It has been shown that super-resolution performance can be improved by utilizing multiple images captured in quick succession \cite{bhat2021deep}. However, this approach is impractical when it comes to HSIs due to the slow acquisition times. Data fusion techniques can be applied to HSIs \cite{xue2021spatial, li2022deep}. However, these approaches rely on the availability of a high-resolution multispectral image of the same scene.

Transformers \cite{vaswani2017attention} are gaining popularity in the vision community and some researchers have utilized them for super-resolution \cite{liang2021swinir}. However, this approach suffers the same problem as the unsupervised GAN methods in that that they require very large amounts of data to be trained. Furthermore, the use of these techniques is also known to be computationally expensive during the inference process.

Given the limited availability of training data makes the use of transformers, modern deep GANs, and CNNs difficult to apply to HSI SR problems. Furthermore, we generally aim to improve the quality of the hyperspectral data before inference tasks, which means that an efficient SR network that can operate in real-time is preferred. The large amount of data to be processed in hyperspectral imaging presents a challenge to using deep networks, so we propose a highly efficient SR neural network structure based on a new paradigm, self-operational neural filtering. 

\subsection{Operational Neural Networks}
\noindent Recent advances in deep learning have resulted in CNNs dominating many computer vision fields, including super-resolution. Part of the reason for their success is their ability to learn complex non-linear operators. However, convolution itself is a linear operation and the non-linear components of the networks are solely provided by the activation functions used after each convolutional layer in the network. This means that CNNs often have to be very deep in order to have the necessary non-linear capacity and diversity to learn the complex function of the learning problem. 

Recently, Operational Neural Networks (ONNs) \cite{kiranyaz2020operational, malik2020fastonn} were proposed to address this issue by incorporating non-linear nodal and pooling functions that replace the sole convolution operation with any non-linear operator, which adds significantly more non-linear components to the network than a traditional CNN. However, these additional non-linear operations are hard coded and thus cannot be changed during training. This means that the functions need to be searched for, which is computationally expensive, and the search space is limited to the function set, which may not contain the optimal function(s). 

The authors of \cite{kiranyaz2020operational} then addressed these limitations by proposing self-organized operational neural networks (Self-ONNs) \cite{kiranyaz2021self} which aim to make the linear filters of a standard CNN non-linear through the use of MacLaurin series expansions, rather than applying hard-coded functions. Such non-linear filters for each kernel element are learnable during training, and thus, eliminate the need for an exhaustive search to find the optimal functions. Furthermore, any function can theoretically be approximated using MacLaurin series expansions, which means that a Self-ONN is not limited to a specified function set, allowing for an enhanced non-linear search space. These improvements mean that Self-ONNs are far more computationally efficient than their standard ONN counterparts, with greater theoretical non-linear capacity than both their ONN and CNN counterparts. This additional complexity comes at the cost of each filter requiring more parameters. However, the network size of a Self-ONN can be much smaller than a CNN to have the same or increased theoretical non-linear capacity, allowing for the overall model to have fewer parameters than a CNN despite each individual filter containing more parameters. In many applications \cite{kiranyaz2021robust, malik2021real, devecioglu2021real, zahid2022global, ince2022blind, malik2021bm3d} Self-ONNs outperformed the deeper and more complex CNNs whilst achieving an elegant computational efficiency.

%In this section we have discussed the difference between ONNs and Self-ONNs to introduce the reader to these new network types. In this work we only make use of Self-ONNs and for the sake of brevity, will refer to Self-ONNs as ONNs.

%However, in our view, the introduction of Self-ONNs has rendered the ONN obsolete, we therefore only use and refer to Self-ONNs in the rest of this paper and for the sake of brevity, will refer to Self-ONNs as ONNs.

%\cite{dian2017hyperspectral}

\section{Methodology}\label{sec:methodology}
\begin{figure}[!t]
    \centering
    \includegraphics[width=0.5\textwidth]{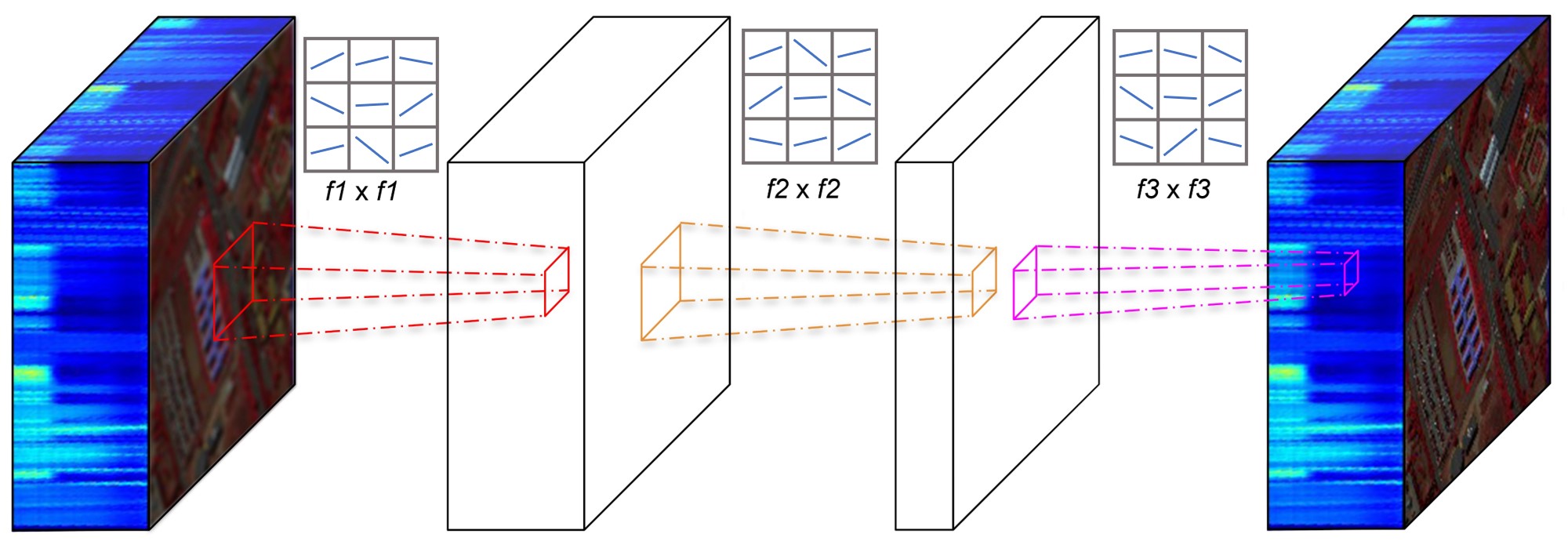}
    \caption{SRCNN model representation consisting of 3 convolutional layers with filter sizes f1 x f1, f2 x f2, and f3 x f3.}
    \label{fig:SRCNN}
\end{figure}

\noindent We take the super-resolution model SRCNN \cite{dong2015image} and modify it for use on hyperspectral images by extending the number of input and output channels of the model from 3 (for RGB images) to the required number for the relevant HSI depending on the number of wavelength bands it contains. SRCNN, shown in Figure \ref{fig:SRCNN}, is a relatively compact model consisting of 3 convolutional layers followed by ReLU activation functions, except for the output layer, where no activation function is used. Although there are many improved variants of SRCNN, we select this model due to its simplicity and wide use. Its simplicity allows us to easily and effectively compare CNN and Self-ONN performance so we can have a high degree of certainty that the performance improvement is solely due to the Self-ONN non-linear filters and not influenced by any other auxiliary network components.
Furthermore, this architecture allows us to examine the effects of incorporating auxiliary components such as residual connections and normalization layers into our Self-ONN models. A shallow model such as SRCNN is also much less prone to overfitting, which is useful for our datasets which are very limited in size.

\begin{figure}[!t]
    \centering
    \includegraphics[width=0.5\textwidth]{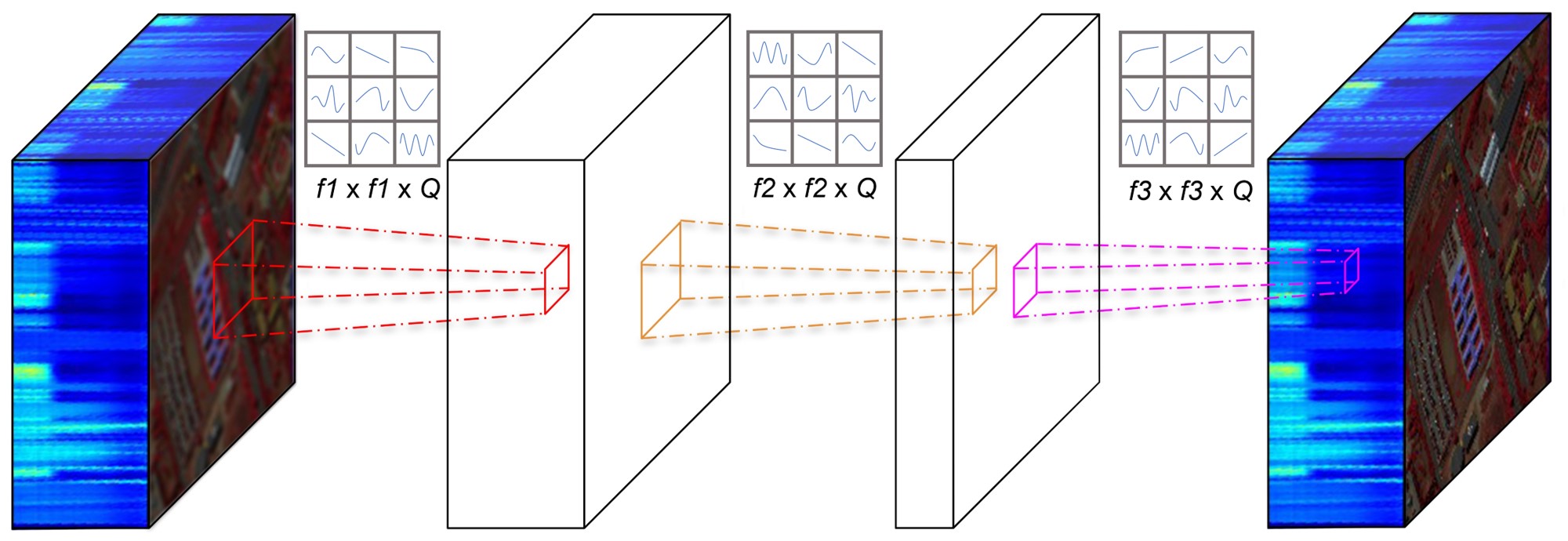}
    \caption{SRONN model representation consisting of 3 self-operational layers with filter sizes (f1 x f1 x Q), (f2 x f2 x Q), and (f3 x f3 x Q). Note, each filter element is a learnable non-linear function, enhancing its theoretical learning capacity over a standard CNN where each filter element is a learnable linear function.}
    \label{fig:SRONN}
\end{figure}

We propose a novel Self-ONN model, SRONN, that shares the same configuration as SRCNN as shown in Figure \ref{fig:SRONN}. A key aspect of Self-ONNs is that data passed between layers must be bounded between $-1$ and $1$ in order to prevent exponentially large values due to the non-linear nature of the model. We, therefore, use hyperbolic tangent (tanh) activation functions after the first and second operational layers in our SRONN model instead of the ReLU activation functions of SRCNN. The tanh activation function is defined in equation \ref{eq:Tanh} and it's output bounds are between $-1$ and $1$, making it an ideal activation function to constrain the data passed between layers to the desired range. 

\begin{equation}
\label{eq:Tanh}
    tanh(x) = \frac{e^{x} - e^{-x}}{e^{x} + e^{-x}}
\end{equation}

\subsection{Parametric Analysis}
\noindent Self-ONNs gain their additional non-linear complexity through the use of MacLaurin series expansions:

\begin{equation}
\label{eq:MacLaurin}
    f(x) = \sum^\infty_{n=0} \frac{f^{(n)}(0)}{n!} x^n
\end{equation}

In practice, the 0th term in the expansion is the bias. Therefore, the 0th term can be disregarded from the filter approximation. The order of the polynomial should be finite in practice so the number of terms is supplied to the network by a parameter $Q$. This makes the expansion for an ONN as follows:

\begin{equation}
\label{eq:ONNMacLaurin}
    f(x) = \sum^Q_{n=1} \frac{f^{(n)}(0)}{n!} x^n
\end{equation}

Note that when the $Q$ value is 1, it is the exact equivalent of a standard convolutional layer. Higher $Q$ values yield more accurate function approximations but at the cost of additional parameters as the $Q$ value directly equates to the multiplication in parameters over a standard convolutional filter. The number of parameters in the convolutional layers of a CNN can be calculated using the following equation:

\begin{equation}
\label{eq:CNNParams}
    \text{\# parameters} = \sum^{l-1}_{l=0} (n_l \times m_l \times f_l + 1) \times f_{l+1}
\end{equation}

where $l$ is the number of layers, $n_l$, $m_l$ is the number of rows and columns in the convolutional filters at layer $l$, $f$ is the number of filters and the constant 1 accounts for the bias for each filter. Note, that on the first layer, i.e. $l=0$, the number of filters from the previous layer ($l\! -\! 1$) is given by the number of channels of the input image. To compute the number of parameters of a Self-ONN, we simply multiply this by $Q$:

\begin{equation}
\label{eq:ONNParams}
    \text{\# parameters} = \sum^{l-1}_{l=0} (n_l \times m_l \times f_l \times Q + 1) \times f_{l+1}
\end{equation}

%We select a $Q$ value of 3 for all our experiments as we found this to be a good balance between sufficient approximation accuracy and the number of parameters. However, this means that each of our SRONN models will have approximately three times as many parameters as the SRCNN models. For a fair comparison, we also propose another Self-ONN model again with the same number of layers as SRCNN, but four times fewer filters per layer. As a result, this model has between 26.5\% and 28.2\% fewer parameters than SRCNN depending on the number of input and output channels required by the target dataset. We name this model small SRONN, or sSRONN.

Our SRONN model will therefore have approximately $Q$ times more parameters than the SRCNN model. To ensure a fair comparison between CNN and Self-ONN, we choose a low $Q$ value. The minimum $Q$ value is 2, as a $Q$ value of 1 is the equivalent of a CNN. However, a $Q$ value of 2 would only add one non-linear term to Eq. (\ref{eq:ONNMacLaurin}), limiting the non-linear function approximation capacity. To enhance this capacity, we use a $Q$ value of 3 in all experiments, which introduces a second non-linear term to Eq. (\ref{eq:ONNMacLaurin}), significantly improving the non-linear function approximation while still keeping the parameter increase relatively low. It is also worth noting that going much beyond this Q value will likely have diminishing performance returns relative to the parameter increase and may even be detrimental to performance due to the increased training difficulty, especially on small datasets. However, a $Q$ value of 3 still means that each SRONN model has around three times more parameters than its equivalent SRCNN model. For a fair comparison, we also propose a Self-ONN model with the same number of layers as SRCNN but with four times fewer filters per layer. This model has between 26.5\% and 28.2\% fewer parameters than SRCNN, depending on the required input and output channels of the dataset. We refer to this model as small SRONN or sSRONN.

To implement a Self-ONN layer in practice a standard convolutional layer can simply be extended by increasing the number of input channels by a factor of $Q$ and passing the input concatenated with the input raised to the power $n$ up to $Q$. The convolutional layer will then apply its weights to all the MacLaurin series terms and perform the required summation of the terms, providing the non-linear learnable MacLaurin series approximation. This practical implementation can be found in the GitHub repository from \cite{malik2020fastonn}. More detailed information about Self-ONNs is presented in Appendix \ref{appendix:Self-ONN}.

\subsection{Normalization and Residual Connections}
\noindent Due to the recent proposal of Self-ONNs \cite{kiranyaz2021self}, techniques commonly applied to CNNs to improve results have been studied little on Self-ONNs. We study the effects of incorporating various normalization layer types into our ONN models after each Tanh activation function including L1, L2, instance \cite{ulyanov2016instance}, and batch \cite{ioffe2015batch} normalization. We also study the effects of adding a residual connection to connect the output of the model directly to the input of the model so that the model learns the residual rather than the direct mapping as performed in \cite{song2018super}. To the best of our knowledge, this is the first work to study the effects of these techniques on Self-ONNs.

\begin{figure}[!t]
    \centering
    \includegraphics[width=0.5\textwidth]{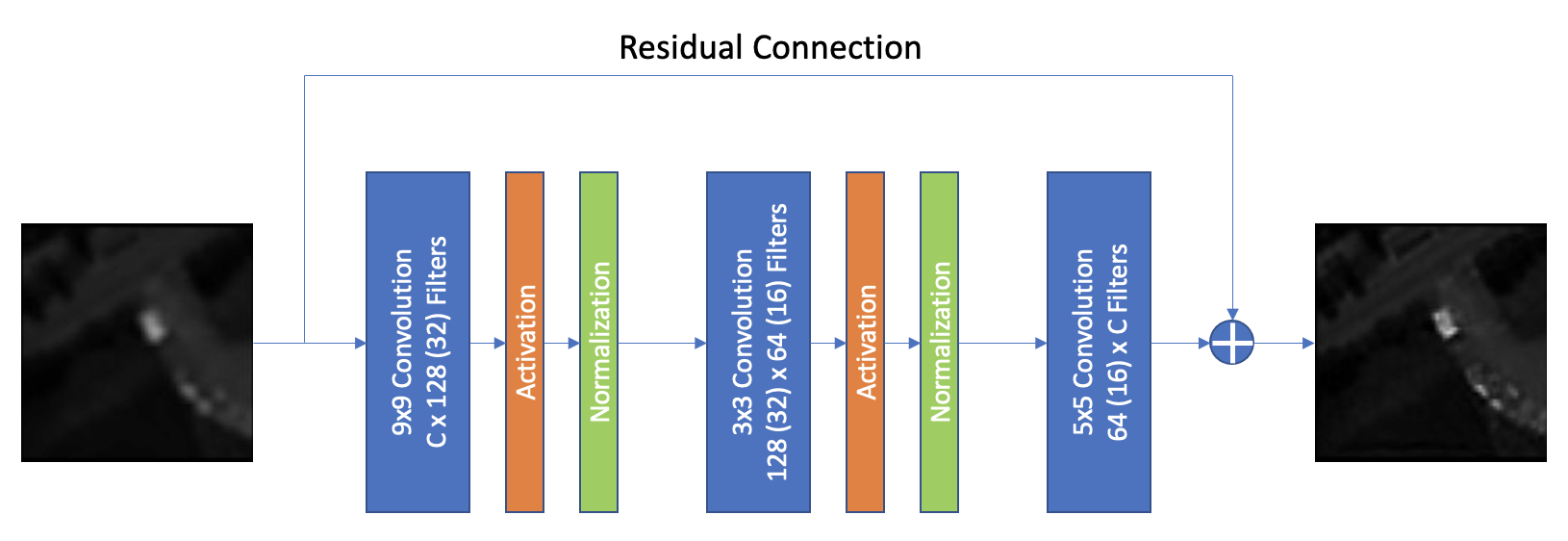}
    \caption{General model architecture. C represents the number of channels in the hyperspectral image. Values in brackets represent the number of filters in the compact sSRONN model. SRCNN and SRONN variants have Cx128, 128x64, and 64xC filters in each respective layer. sSRONN variant has Cx32, 32x16, and 16xC filters in each respective layer. The normalization type depends on the experiment and in some experiments, there is no normalization, in which case the normalization layers are skipped. The residual connection is also removed in experiments where it is not applied.}
    \label{fig:Model}
\end{figure}

The proposed Self-ONN model is illustrated in Figure \ref{fig:Model}.

\section{Results}\label{sec:results}

\begin{figure*}[!t]
    \centering
    \includegraphics[width=1\textwidth]{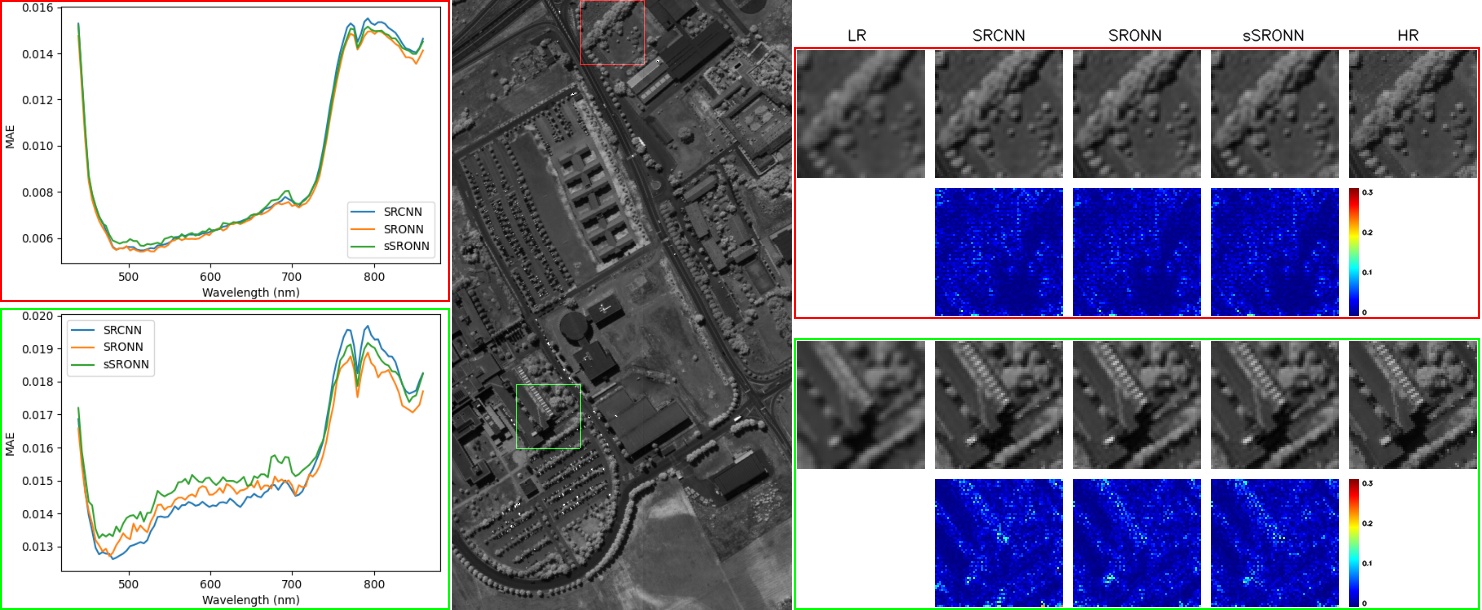}
    \caption{Output of models with no residual connection or normalization on the Pavia University dataset. The mean absolute error between the predicted and true spectra across the patch is shown on the left. Slice 80 of the original HSI is shown in the middle. LR, predictions, HR and the absolute difference between prediction and HR are shown on the right.}
    \label{fig:regResults}
\end{figure*}

\noindent We first compare the SRCNN models against the SRONN and sSRONN models without normalization for a fair comparison. The results can be seen in Table \ref{tab:regResults} and example outputs on the Pavia University dataset from the models with and without residual connections can be seen in Figure \ref{fig:regResults} and Figure \ref{fig:residualResults} respectively. True super-resolution outputs, i.e. where there is no target image and super-resolution is performed on the original data (no downsampling), on the Pavia University dataset can be seen in Figure \ref{fig:trueSRResults} and Figure \ref{fig:residualTrueSRResults}.

We only apply normalization to the Self-ONN models, since normalization has been widely studied in CNNs. We present the results from adding various normalization types to the Self-ONN models in separate tables for each dataset. Results for the Cuprite dataset are shown in Table~\ref{tab:cupriteNormResults}, Pavia University in Table~\ref{tab:paviaNormResults}, Salinas in Table~\ref{tab:salinasNormResults} and Urban in Table~\ref{tab:urbanNormResults} within Appendix \ref{appendix:norm}.

For the three training iterations of each model on each dataset, we report only the results from the best iteration in each table of results.

\subsection{Datasets}
\noindent We evaluate our models on four different HSI datasets: Pavia University; Salinas; Cuprite; Urban. Details for each dataset \cite{HS-RSS, rslab} can be seen in Table~\ref{tab:dataInfo}.

%Each model was trained on the Pavia University, Urban, Cuprite, and Salinas HSI datasets which were preprocessed by min-max normalization and then divided into 64x64 pixel tiles, maintaining the entire wavelength spectrum. 

%Salinas. Collected from 224-band AVIRIS sensor over Salinas Vally, California. 3.7 meters per pixel spatial resolution. 512x217 image size (I think). 20 water absorption bands removed [108-112], [154-167], 224.

%Cuprite. Aquired by AVIRIS sensor over Las Vegas, Nevada. 224 bands (307nm to 2480nm), noise channels 1-2 and 221-224 and water absorbtion channels 104-113 and 148-167 removed, 188 channels remaining. 250x190 pixels.

%Urban (unsure what sensor) 210 bands (400nm to 2500nm), 307x307 pixels, 2m resolution. Channels 1-4, 76, 87, 101-111, 136-153 and 198-210 removed, 162 remaining.

%Pavia University. Acquired by ROSIS sensor over Pavia, Italy. 103 bands, 610x610 pixels, 1.3m resolution. 

\begin{table}[!ht]
    \caption{Dataset Information}
    \label{tab:dataInfo}
    \centering
    \begin{tabular}{ | c | c | c | c | }
    	\hline
    	Dataset & Image Dimensions & Channels & Resolution \\ \hline
	Pavia University & $610 \times 340$ & $103$ & $1.3m$ \\ 
	Salinas & $512 \times 217$ & $204$ & $3.7m$ \\ 
	Cuprite & $512 \times 614$ & $224$ & - \\ 
	Urban & $307 \times 307$ & $210$ & $2m$ \\ \hline
	
    \end{tabular}
\end{table}

We use the standard approach to generating a low-resolution image pair from a given high-resolution target image by using Eq. (\ref{eq:LRequation}):
\begin{equation}
\label{eq:LRequation}
    I_{LR} = (I_{HR}*k)\downarrow_s+\: n
\end{equation}

where $k \in R^2$ is a 2D degradation kernel, * is a spatial convolution, $\downarrow_s$ is a decimation operation with a stride s, and n is a noise term. We use Gaussian blur with a sigma value of 0.8943 for k as is done in \cite{wang2020frequency}, 2$\times$ subsampling for $\downarrow_s$. We do not add any noise so the parameter n is ignored. Each generated LR tile was then bilinearly interpolated back up to the size of the original tile so the model could perform super-resolution by recovering the information at the desired output resolution. The model would then be trained with the LR tile as input and the original HR tile as the target. We select a scale factor of 2$\times$ as the datasets we are using are very small in size, making it infeasible to go beyond this scale factor.

%and Gaussian noise with variance $5 \times 10^{-5}$ for n

Each dataset was preprocessed with min-max normalization and then divided into 64x64 pixel tiles, maintaining the entire wavelength spectrum. We utilize 70\% of the tiles for training, 15\% for validation and reserve 15\% for testing.

\subsection{Training Details}
\noindent Each model was trained for 50000 epochs to guarantee network convergence, and the weights from the epoch which produced the highest SSIM validation score were used for testing. We use the Adam optimizer \cite{kingma2014adam} with default parameters except for the learning rate. Each model was initially trained with a learning rate of $10^{-4}$ which was decreased by a factor of 10 at epochs 5000 and 40000. Two following runs were then completed where the starting learning rate and the epoch milestones - where the learning rate was decreased by a factor of 10 - were manually adjusted in an attempt to improve the performance. We use mean squared error as our loss function. We initialize our models' weights with a normal distribution with a gain of 0.02. All training LR tiles are fed to the model in a single batch on each epoch. For all experiments, the entire training dataset was forward propagated through the model at once so there was no need to adjust the batch size.

%We trained each model combination 3 times, i.e. the combination of the base model, the normalization type used, and whether or not there was a residual connection. The first iteration uses the default parameters of $10^{-4}$ learning rate, and the learning rate is decreased by a factor of 10 on epochs 5000 and 40000. For the remaining two iterations, the parameters were manually adjusted in an attempt to improve the performance from the default first iteration. 

\begin{table*}
    \caption{Results from Standard Models with no Normalization.}
    \label{tab:regResults}
    \centering
    \begin{tabular}{|c|c|c|c|c|c|c|c|c|}
    \hline
    Dataset & Model & Residual & \# parameters & lr & lr steps & PSNR & SSIM & SAM \\ \hline\hline
    \multirow{6}{5em}{Cuprite} & SRCNN & \multirow{3}{3em}{no} & 2754976 & $10^{-4}$ & 100k & 27.799 & \textbf{0.9766} & 10.136 \\ \cline{2-2}\cline{4-9}
    ~ & SRONN & ~ & 8264096 & $10^{-4}$ & 2.5k & \textbf{27.882} & 0.9743 & \textbf{10.044} \\ \cline{2-2}\cline{4-9}
    ~ & sSRONN & ~ & \textbf{2024720} & $10^{-4}$ & 15k & 27.863 & 0.9746 & 10.061 \\ \cline{2-9}\cline{2-9}
    ~ & SRCNN & \multirow{3}{3em}{yes} & 2754976 & $10^{-4}$ & 5k, 40k & 27.783 & 0.9731 & 10.118 \\ \cline{2-2}\cline{4-9}
    ~ & SRONN & ~ & 8264096 & $10^{-4}$ & 2.5k & 27.927 & 0.9774 & 9.993 \\ \cline{2-2}\cline{4-9}
    ~ & sSRONN & ~ & \textbf{2024720} & $10^{-4}$ & 2.5k & \textbf{27.959} & \textbf{0.9775} & \textbf{9.961} \\ \hline\hline
        
    \multirow{6}{5em}{Pavia \par University} & SRCNN & \multirow{3}{3em}{no} & 1306727 & $10^{-4}$ & 5k, 40k & 35.396 & 0.977 & 4.346 \\ \cline{2-2}\cline{4-9}
    ~ & SRONN & ~ & 3919591 & $10^{-4}$ & 2.5k & \textbf{35.857} & \textbf{0.9775} & \textbf{4.209} \\ \cline{2-2}\cline{4-9}
    ~ & sSRONN & ~ & \textbf{938503} & $10^{-4}$ & 50k & 35.693 & 0.9768 & 4.606 \\ \cline{2-9}\cline{2-9}
    ~ & SRCNN & \multirow{3}{3em}{yes} & 1306727 & $10^{-4}$ & 2.5k, 10k, 30k & 35.597 & 0.9768 & 4.388 \\ \cline{2-2}\cline{4-9}
    ~ & SRONN & ~ & 3919591 & $10^{-4}$ & 5k, 40k & 35.914 & \textbf{0.9783} & 4.056 \\ \cline{2-2}\cline{4-9}
    ~ & sSRONN & ~ & \textbf{938503} & $10^{-4}$ & 5k, 40k & \textbf{35.926} & 0.9782 & \textbf{4.033} \\ \hline\hline
    
    \multirow{6}{5em}{Salinas} & SRCNN & \multirow{3}{3em}{no} & 2515596 & $10^{-4}$ & 5k & \textbf{44.074} & \textbf{0.9943} & \textbf{1.462} \\ \cline{2-2}\cline{4-9}
    ~ & SRONN & ~ & 7545996 & $10^{-4}$ & 2.5k & 43.941 & 0.994 & 1.549 \\ \cline{2-2}\cline{4-9}
    ~ & sSRONN & ~ & \textbf{1845180} & $10^{-4}$ & 5k, 40k & 43.558 & 0.9937 & 1.622 \\ \cline{2-9}\cline{2-9}
    ~ & SRCNN & \multirow{3}{3em}{yes} & 2515596 & $10^{-4}$ & 5k, 40k & 44.025 & 0.9941 & 1.517 \\ \cline{2-2}\cline{4-9}
    ~ & SRONN & ~ & 7545996 & $10^{-4}$ & 10k & 44.223 & 0.9944 & 1.461 \\ \cline{2-2}\cline{4-9}
    ~ & sSRONN & ~ & \textbf{1845180} & $10^{-4}$ & 4.5k, 30k & \textbf{44.286} & \textbf{0.9945} & \textbf{1.412} \\ \hline\hline
        
    \multirow{3}{5em}{Urban} & SRCNN & \multirow{3}{3em}{no} & 2587410 & $10^{-4}$ & 5k, 40k & 25.231 & 0.8878 & 14.811 \\ \cline{2-2}\cline{4-9}
    ~ & SRONN & ~ & 7761426 & $10^{-4}$ & 5k, 40k & \textbf{25.941} & \textbf{0.8935} & \textbf{13.94} \\ \cline{2-2}\cline{4-9}
    ~ & sSRONN & ~ & \textbf{1899042} & $10^{-4}$ & 3k & 25.818 & 0.8912 & 14.22 \\ \cline{2-9}\cline{2-9}
    ~ & SRCNN & \multirow{3}{3em}{yes} & 2587410 & $10^{-5}$ & 5k, 40k & 25.872 & 0.8916 & 13.958 \\ \cline{2-2}\cline{4-9}
    ~ & SRONN & ~ & 7761426 & $10^{-4}$ & 2k & 25.892 & \textbf{0.8999} & \textbf{13.613} \\ \cline{2-2}\cline{4-9}
    ~ & sSRONN & ~ & \textbf{1899042} & $10^{-4}$ & 4k & \textbf{26.065} & 0.8963 & 13.681 \\ \hline
    \end{tabular}
\end{table*}

\begin{figure*}[!t]
    \centering
    \includegraphics[width=1\textwidth]{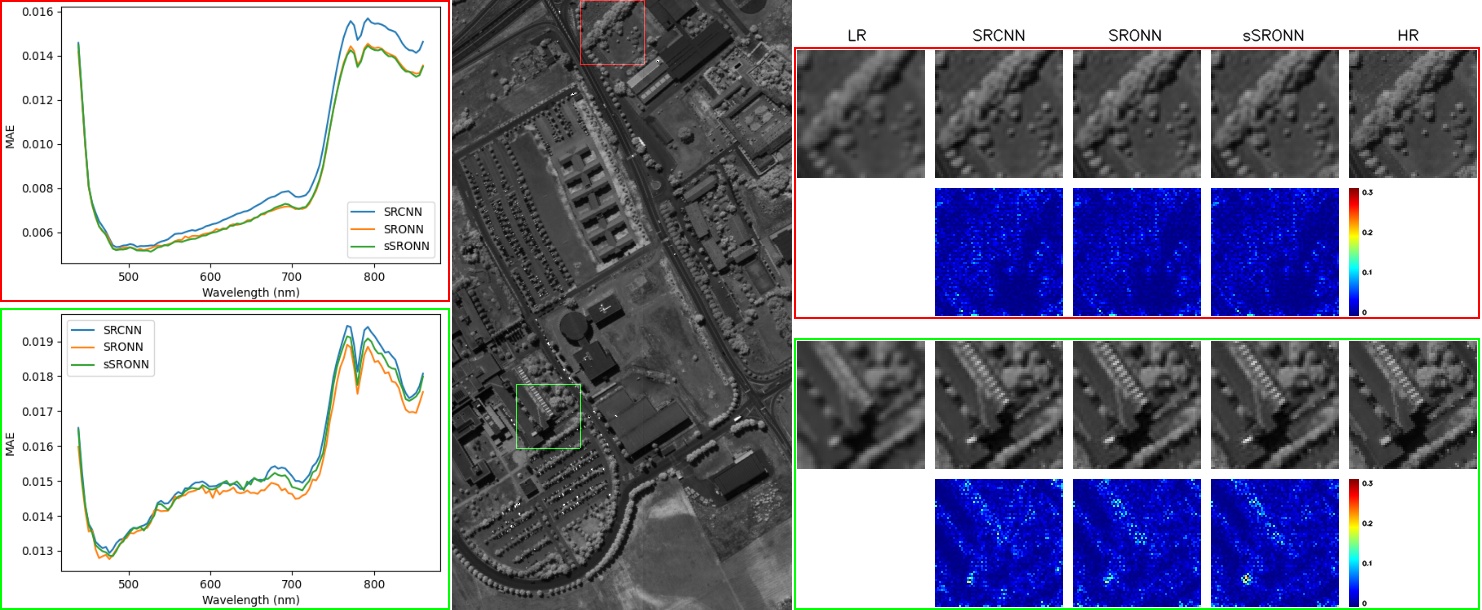}
    \caption{Output of models with a residual connection on the Pavia University dataset. The mean absolute error between the predicted and true spectra across the patch is shown on the left. Slice 80 of the original HSI is shown in the middle. LR, predictions, HR and the absolute difference between prediction and HR are shown on the right.}
    \label{fig:residualResults}
\end{figure*}

\begin{figure*}[!t]
    \centering
    \includegraphics[width=1\textwidth]{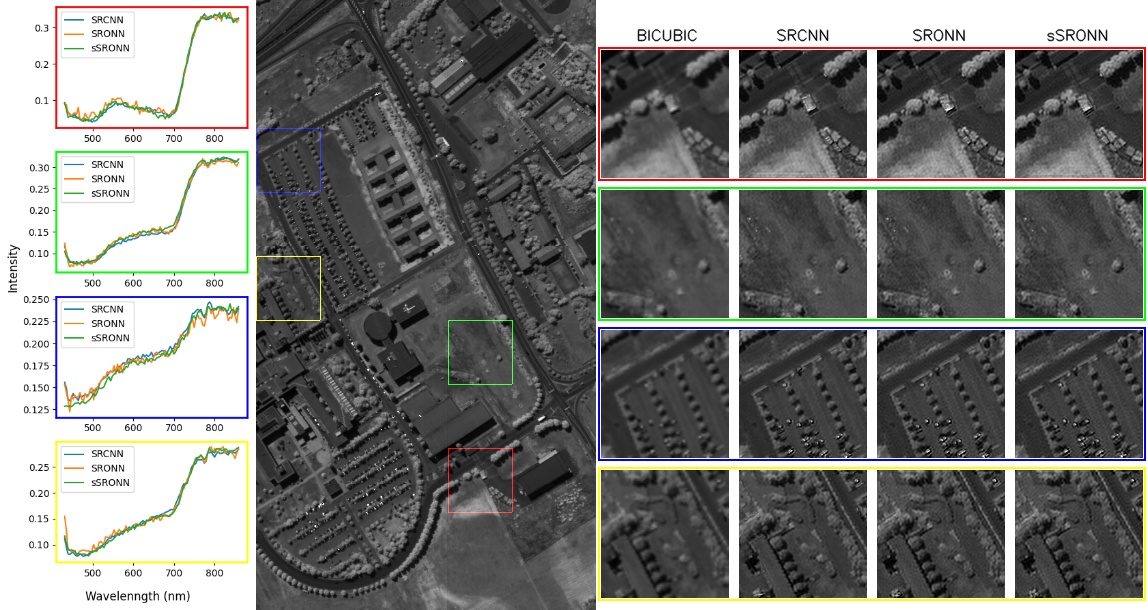}
    \caption{True super-resolution output of models with no residual connection on slice 80 of the Pavia University dataset. Spectral plots of the center pixel of each coloured image patch for each model output are shown on the left. The original HSI is shown in the middle. Test tiles bilinearly interpolated up to 2x their original size and super-resolution results on the interpolated tiles are shown on the right.}
    \label{fig:trueSRResults}
\end{figure*}

\begin{figure*}[!t]
    \centering
    \includegraphics[width=1\textwidth]{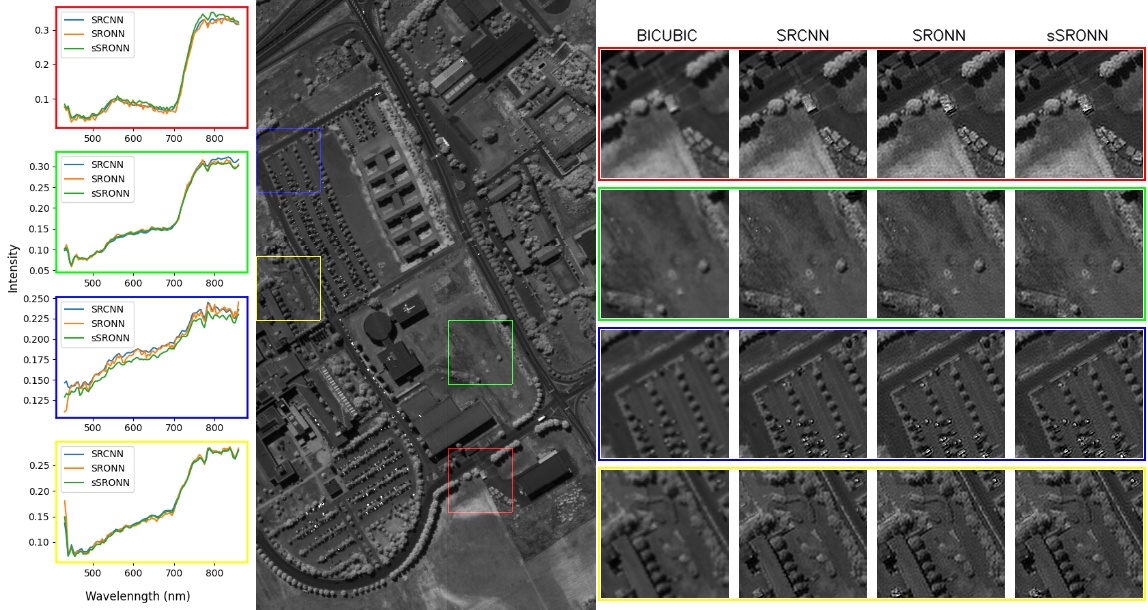}
    \caption{True super-resolution output of models with a residual connection on slice 80 of the Pavia University dataset. Spectral plots of the center pixel of each coloured image patch for each model output are shown on the left. The original HSI is shown in the middle. Test tiles bilinearly interpolated up to 2x their original size and super-resolution results on the interpolated tiles are shown on the right.}
    \label{fig:residualTrueSRResults}
\end{figure*}

\section{Discussion}\label{sec:discussion}
%\noindent The results from Table~\ref{tab:regResults} show that the base SRONN models with no residual connection tend to offer a slight improvement over the SRCNN model with no residual connection, but this is not always the case. For example, on the Salinas dataset, the SRCNN model outperformed the SRONN models across all metrics. We hypothesise that this is due to the Self-ONN models having a more complex search space to navigate and optimise due to the non-linear nature of the filters, and therefore have more difficulty converging than the simpler SRCNN model.

\noindent The results from Table~\ref{tab:regResults} reveal that the base SRONN models without a residual connection generally offer a slight improvement over the SRCNN model that also lacks a residual connection. However, an exception to this trend occurs specifically in the Salinas dataset, where the SRCNN model without a residual connection outperformed the corresponding SRONN models across all metrics. It is essential to note that this outperformance is confined only to the Salinas dataset and is not representative of the overall trend observed across the other three experimental datasets. We hypothesise that this may be due to the Self-ONN models having a more complex search space to navigate and optimise, owing to the non-linear nature of the filters, thus causing more difficulty in converging compared to the simpler SRCNN model. %Furthermore, when a residual connection is incorporated into each model, both SRONN models consistently and invariably outperform the SRCNN model, reinforcing the fact that SRCNN's superior performance is restricted solely to one specific condition on a single dataset.

\subsection{Effects of Residual Connections}
\noindent The results from Table~\ref{tab:regResults} show that adding a residual connection provides significant improvement to both Self-ONN models, resulting in both the SRONN and sSRONN models outperforming the SRCNN models across all metrics on all datasets. The addition of a residual connection improved all metrics across all datasets for both sSRONN and SRONN except for PSNR on the Urban dataset for the SRONN model where a slight decrease was observed. Furthermore, the addition of a residual connection greatly increased convergence time which can be seen in the model training loss and validation metric plots we have included in Appendix \ref{appendix:trainPlots}. The residual connection has a lesser impact on the results of SRCNN, only offering improvement in some cases, which is likely due to the model not being complex enough to see any consistent performance improvement from a residual connection. The residual connection performance improvement on the spectra can be very clearly observed in the mean absolute error spectral plots in Figures 4 \& 5. In Figure 4, the SRONN model tends to have better spectral reconstruction at the higher wavelengths while the SRCNN model is usually better at the lower wavelengths, but from that plot, it is visually difficult to say which is better overall except that they are both better than the sSRONN model. However, when a residual connection is added, the mean absolute error spectral plots in Figure 5 quite conclusively show that both ONN models provide superior spectral reconstruction than the SRCNN model.

The improvement seen in the performance and convergence times of our Self-ONN models when a residual connection is added supports our convergence hypothesis. It could also be indicative that Self-ONNs may suffer more from vanishing gradients than CNNs. Interestingly, the sSRONN model generally saw greater performance improvements from the addition of a residual connection than the SRONN model, which is counterintuitive as the sSRONN optimization search space is significantly smaller than the search space of the SRONN model. One explanation for this could be that the sSRONN model might be slightly under-parameterized for direct image-to-image mapping. However, it may have sufficient parameters to learn the residual, resulting in a bigger performance improvement when the residual connection is added to the model. The larger SRONN model, which may be well-parameterised for image-to-image mapping but slightly over-parameterised for residual learning, does not see as much of a performance improvement as the smaller sSRONN model.

Since both SRONN and sSRONN outperform SRCNN, this demonstrates the power of the non-linear filters over the standard linear convolutional filters. The non-linear filters provide the operational layer with an enhanced ability to produce sharper edges and thus produce sharper contrast between pixels resulting in a more detailed output image, which is evident in the resulting images shown in Figure ~\ref{fig:residualResults} and Figure ~\ref{fig:residualTrueSRResults}.

\subsection{Effects of Normalization}
\noindent Our results in Tables~\ref{tab:cupriteNormResults},~\ref{tab:paviaNormResults},~\ref{tab:salinasNormResults}, and~\ref{tab:urbanNormResults} show the effects of incorporating normalization layers into our SRONN and sSRONN models are largely varied and highly dataset dependent. It appears that normalization has a greater impact on the datasets with larger spatial dimensions. We found L2 normalization to be the most effective, providing a slight performance boost to the SRONN model across all metrics on the Cuprite, Pavia University and Urban datasets while boosting the SAM on the Salinas dataset. For the sSRONN model, the performance improvement from adding L2 normalization is less significant, providing only a performance boost to SSIM and SAM on the Cuprite dataset, PSNR on the Pavia University dataset and SSIM on the Urban dataset. No performance improvement was provided by using L2 normalization over no normalization on the Salinas dataset.

Our results show that normalization is generally more effective when utilized in conjunction with a residual connection. This is likely due to the fact the normalization layers will normalize the data around a zero mean which makes it more difficult for the models without a residual connection to map the zero mean feature maps to the true mean of the output. However, when a residual connection is introduced, the model learns the residual between the input and the target, which should have a mean near zero. Therefore, normalization may offer a greater benefit in this scenario as it assists the model in transforming the data to the target mean, rather than moving it away from the target mean.

Interestingly, we found instance normalization to be especially detrimental to all results. This could be because instance normalization normalizes each channel individually which may have an adverse effect on the channel dependencies.

\section{Conclusion}\label{sec:conclusion}
\noindent We show that Self-ONNs outperform equivalent well known CNNs in the task of HSI SR, even when the Self-ONN models have a lower number of parameters than the CNNs. The Self-ONN results produced sharper images and contained more detail which is likely a direct result of the enhanced non-linear filters. 

We found that adding a residual connection to our SRONN and sSRONN models provided a significant performance improvement and greatly increased convergence times. We hypothesize that Self-ONNs suffer more from the vanishing gradient problem than CNNs due to their more complex search spaces and thus the residual connection helps mitigate this issue, even in relatively shallow models.

We examined the effects of adding a residual connection and various normalization layers to our ONN models. Our results show that L2 normalization layers in ONNs can offer a moderate performance improvement when used in conjunction with a residual connection, but the benefit of normalization appears to be highly dependent on the dataset. 

We show that the superior non-linear capabilities of ONNs compared to CNNs allow for sharper and more detailed HSI SR results. This indicates that Self-ONNs can outperform CNN models in such image-to-image mapping tasks. Finding the best Self-ONN models with the right hyperparameters will be the topic of our future work.

\section{Acknowledgments}
\noindent This work was supported by the Engineering and Physical Sciences Research Council [grant number EP/T517938/1] and Peacock Technology Limited.

\bibliographystyle{IEEEtran}
\bibliography{HSI-ONN-SR.bib}

\begin{thebibliography}{10}
\providecommand{\url}[1]{#1}
\csname url@samestyle\endcsname
\providecommand{\newblock}{\relax}
\providecommand{\bibinfo}[2]{#2}
\providecommand{\BIBentrySTDinterwordspacing}{\spaceskip=0pt\relax}
\providecommand{\BIBentryALTinterwordstretchfactor}{4}
\providecommand{\BIBentryALTinterwordspacing}{\spaceskip=\fontdimen2\font plus
\BIBentryALTinterwordstretchfactor\fontdimen3\font minus
  \fontdimen4\font\relax}
\providecommand{\BIBforeignlanguage}[2]{{%
\expandafter\ifx\csname l@#1\endcsname\relax
\typeout{** WARNING: IEEEtran.bst: No hyphenation pattern has been}%
\typeout{** loaded for the language `#1'. Using the pattern for}%
\typeout{** the default language instead.}%
\else
\language=\csname l@#1\endcsname
\fi
#2}}
\providecommand{\BIBdecl}{\relax}
\BIBdecl

\bibitem{HS-RS}
``Hyperspectral remote sensing,'' The University of Texas at Austin
  \url{http://www.csr.utexas.edu/projects/rs/hrs/hyper.html} (accessed Apr. 5,
  2022).

\bibitem{villafranca2012limitations}
A.~G. Villafranca, J.~Corbera, F.~Mart{\'\i}n, and J.~F. March{\'a}n,
  ``Limitations of hyperspectral earth observation on small satellites,''
  \emph{Journal of Small Satellites}, vol.~1, no.~1, pp. 19--29, 2012.

\bibitem{brady2009optical}
D.~J. Brady, \emph{Optical imaging and spectroscopy}.\hskip 1em plus 0.5em
  minus 0.4em\relax John Wiley \& Sons, 2009.

\bibitem{wang2021real}
X.~Wang, L.~Xie, C.~Dong, and Y.~Shan, ``Real-esrgan: Training real-world blind
  super-resolution with pure synthetic data,'' in \emph{Proceedings of the
  IEEE/CVF International Conference on Computer Vision}, 2021, pp. 1905--1914.

\bibitem{wang2022semi}
L.~Wang and K.-J. Yoon, ``Semi-supervised student-teacher learning for single
  image super-resolution,'' \emph{Pattern Recognition}, vol. 121, p. 108206,
  2022.

\bibitem{dong2015image}
C.~Dong, C.~C. Loy, K.~He, and X.~Tang, ``Image super-resolution using deep
  convolutional networks,'' \emph{IEEE transactions on pattern analysis and
  machine intelligence}, vol.~38, no.~2, pp. 295--307, 2015.

\bibitem{singh2022semi}
A.~Singh and P.~Rai, ``Semi-supervised super-resolution,'' \emph{arXiv preprint
  arXiv:2204.08192}, 2022.

\bibitem{kiranyaz2020operational}
S.~Kiranyaz, T.~Ince, A.~Iosifidis, and M.~Gabbouj, ``Operational neural
  networks,'' \emph{Neural Computing and Applications}, vol.~32, no.~11, pp.
  6645--6668, 2020.

\bibitem{malik2020fastonn}
J.~Malik, S.~Kiranyaz, and M.~Gabbouj, ``Fastonn--python based open-source gpu
  implementation for operational neural networks,'' \emph{arXiv preprint
  arXiv:2006.02267}, 2020.

\bibitem{kiranyaz2021self}
S.~Kiranyaz, J.~Malik, H.~B. Abdallah, T.~Ince, A.~Iosifidis, and M.~Gabbouj,
  ``Self-organized operational neural networks with generative neurons,''
  \emph{Neural Networks}, vol. 140, pp. 294--308, 2021.

\bibitem{rslab}
``Remote sensing datasets,'' Remote Sensing Laboratory School of Surveying and
  Geospatial Engineering \url{https://rslab.ut.ac.ir/data} (accessed Mar. 4,
  2022).

\bibitem{HS-RSS}
``Hyperspectral remote sensing scenes,'' University of the Basque Country
  \url{https://www.ehu.eus/ccwintco/index.php/Hyperspectral_Remote_Sensing_Scenes}
  (accessed Mar. 16, 2022).

\bibitem{lim2017enhanced}
B.~Lim, S.~Son, H.~Kim, S.~Nah, and K.~Mu~Lee, ``Enhanced deep residual
  networks for single image super-resolution,'' in \emph{Proceedings of the
  IEEE conference on computer vision and pattern recognition workshops}, 2017,
  pp. 136--144.

\bibitem{shi2016real}
W.~Shi, J.~Caballero, F.~Husz{\'a}r, J.~Totz, A.~P. Aitken, R.~Bishop,
  D.~Rueckert, and Z.~Wang, ``Real-time single image and video super-resolution
  using an efficient sub-pixel convolutional neural network,'' in
  \emph{Proceedings of the IEEE conference on computer vision and pattern
  recognition}, 2016, pp. 1874--1883.

\bibitem{ledig2017photo}
C.~Ledig, L.~Theis, F.~Husz{\'a}r, J.~Caballero, A.~Cunningham, A.~Acosta,
  A.~Aitken, A.~Tejani, J.~Totz, Z.~Wang \emph{et~al.}, ``Photo-realistic
  single image super-resolution using a generative adversarial network,'' in
  \emph{Proceedings of the IEEE conference on computer vision and pattern
  recognition}, 2017, pp. 4681--4690.

\bibitem{yamac2021kernelnet}
M.~Yamac, B.~Ataman, and A.~Nawaz, ``Kernelnet: A blind super-resolution kernel
  estimation network,'' in \emph{Proceedings of the IEEE/CVF Conference on
  Computer Vision and Pattern Recognition}, 2021, pp. 453--462.

\bibitem{kim2016accurate}
J.~Kim, J.~K. Lee, and K.~M. Lee, ``Accurate image super-resolution using very
  deep convolutional networks,'' in \emph{Proceedings of the IEEE conference on
  computer vision and pattern recognition}, 2016, pp. 1646--1654.

\bibitem{arun2020cnn}
P.~V. Arun, K.~M. Buddhiraju, A.~Porwal, and J.~Chanussot, ``Cnn-based
  super-resolution of hyperspectral images,'' \emph{IEEE Transactions on
  Geoscience and Remote Sensing}, vol.~58, no.~9, pp. 6106--6121, 2020.

\bibitem{zheng2018multi}
K.~Zheng, L.~Gao, B.~Zhang, and X.~Cui, ``Multi-losses function based
  convolution neural network for single hyperspectral image super-resolution,''
  in \emph{2018 Fifth International Workshop on Earth Observation and Remote
  Sensing Applications (EORSA)}.\hskip 1em plus 0.5em minus 0.4em\relax IEEE,
  2018, pp. 1--4.

\bibitem{song2018super}
T.-A. Song, S.~R. Chowdhury, K.~Kim, K.~Gong, G.~El~Fakhri, Q.~Li, and
  J.~Dutta, ``Super-resolution pet using a very deep convolutional neural
  network,'' in \emph{2018 IEEE Nuclear Science Symposium and Medical Imaging
  Conference Proceedings (NSS/MIC)}.\hskip 1em plus 0.5em minus 0.4em\relax
  IEEE, 2018, pp. 1--2.

\bibitem{zhao2016loss}
H.~Zhao, O.~Gallo, I.~Frosio, and J.~Kautz, ``Loss functions for image
  restoration with neural networks,'' \emph{IEEE Transactions on computational
  imaging}, vol.~3, no.~1, pp. 47--57, 2016.

\bibitem{goodfellow2014generative}
I.~Goodfellow, J.~Pouget-Abadie, M.~Mirza, B.~Xu, D.~Warde-Farley, S.~Ozair,
  A.~Courville, and Y.~Bengio, ``Generative adversarial nets,'' \emph{Advances
  in neural information processing systems}, vol.~27, 2014.

\bibitem{ignatov2018wespe}
A.~Ignatov, N.~Kobyshev, R.~Timofte, K.~Vanhoey, and L.~Van~Gool, ``Wespe:
  weakly supervised photo enhancer for digital cameras,'' in \emph{Proceedings
  of the IEEE Conference on Computer Vision and Pattern Recognition Workshops},
  2018, pp. 691--700.

\bibitem{kim2020unsupervised}
G.~Kim, J.~Park, K.~Lee, J.~Lee, J.~Min, B.~Lee, D.~K. Han, and H.~Ko,
  ``Unsupervised real-world super resolution with cycle generative adversarial
  network and domain discriminator,'' in \emph{Proceedings of the IEEE/CVF
  Conference on Computer Vision and Pattern Recognition Workshops}, 2020, pp.
  456--457.

\bibitem{huang2019hyperspectral}
Q.~Huang, W.~Li, T.~Hu, and R.~Tao, ``Hyperspectral image super-resolution
  using generative adversarial network and residual learning,'' in \emph{ICASSP
  2019-2019 IEEE International Conference on Acoustics, Speech and Signal
  Processing (ICASSP)}.\hskip 1em plus 0.5em minus 0.4em\relax IEEE, 2019, pp.
  3012--3016.

\bibitem{bell2019blind}
S.~Bell-Kligler, A.~Shocher, and M.~Irani, ``Blind super-resolution kernel
  estimation using an internal-gan,'' \emph{Advances in Neural Information
  Processing Systems}, vol.~32, 2019.

\bibitem{wang2004image}
Z.~Wang, A.~C. Bovik, H.~R. Sheikh, and E.~P. Simoncelli, ``Image quality
  assessment: from error visibility to structural similarity,'' \emph{IEEE
  transactions on image processing}, vol.~13, no.~4, pp. 600--612, 2004.

\bibitem{karras2020training}
T.~Karras, M.~Aittala, J.~Hellsten, S.~Laine, J.~Lehtinen, and T.~Aila,
  ``Training generative adversarial networks with limited data,''
  \emph{Advances in Neural Information Processing Systems}, vol.~33, pp.
  12\,104--12\,114, 2020.

\bibitem{johnson2016perceptual}
J.~Johnson, A.~Alahi, and L.~Fei-Fei, ``Perceptual losses for real-time style
  transfer and super-resolution,'' in \emph{European conference on computer
  vision}.\hskip 1em plus 0.5em minus 0.4em\relax Springer, 2016, pp. 694--711.

\bibitem{bhat2021deep}
G.~Bhat, M.~Danelljan, L.~Van~Gool, and R.~Timofte, ``Deep burst
  super-resolution,'' in \emph{Proceedings of the IEEE/CVF Conference on
  Computer Vision and Pattern Recognition}, 2021, pp. 9209--9218.

\bibitem{xue2021spatial}
J.~Xue, Y.-Q. Zhao, Y.~Bu, W.~Liao, J.~C.-W. Chan, and W.~Philips,
  ``Spatial-spectral structured sparse low-rank representation for
  hyperspectral image super-resolution,'' \emph{IEEE Transactions on Image
  Processing}, vol.~30, pp. 3084--3097, 2021.

\bibitem{li2022deep}
J.~Li, K.~Zheng, J.~Yao, L.~Gao, and D.~Hong, ``Deep unsupervised blind
  hyperspectral and multispectral data fusion,'' \emph{IEEE Geoscience and
  Remote Sensing Letters}, vol.~19, pp. 1--5, 2022.

\bibitem{vaswani2017attention}
A.~Vaswani, N.~Shazeer, N.~Parmar, J.~Uszkoreit, L.~Jones, A.~N. Gomez,
  {\L}.~Kaiser, and I.~Polosukhin, ``Attention is all you need,''
  \emph{Advances in neural information processing systems}, vol.~30, 2017.

\bibitem{liang2021swinir}
J.~Liang, J.~Cao, G.~Sun, K.~Zhang, L.~Van~Gool, and R.~Timofte, ``Swinir:
  Image restoration using swin transformer,'' in \emph{Proceedings of the
  IEEE/CVF International Conference on Computer Vision}, 2021, pp. 1833--1844.

\bibitem{kiranyaz2021robust}
S.~Kiranyaz, J.~Malik, M.~U. Zahid, T.~Ince, M.~Chowdhury, A.~Khandakar,
  A.~Tahir, and M.~Gabbouj, ``Robust peak detection for holter ecgs by
  self-organized operational neural networks,'' \emph{arXiv preprint
  arXiv:2110.02381}, 2021.

\bibitem{malik2021real}
J.~Malik, O.~C. Devecioglu, S.~Kiranyaz, T.~Ince, and M.~Gabbouj, ``Real-time
  patient-specific ecg classification by 1d self-operational neural networks,''
  \emph{IEEE Transactions on Biomedical Engineering}, vol.~69, no.~5, pp.
  1788--1801, 2021.

\bibitem{devecioglu2021real}
O.~C. Devecioglu, J.~Malik, T.~Ince, S.~Kiranyaz, E.~Atalay, and M.~Gabbouj,
  ``Real-time glaucoma detection from digital fundus images using self-onns,''
  \emph{IEEE Access}, vol.~9, pp. 140\,031--140\,041, 2021.

\bibitem{zahid2022global}
M.~U. Zahid, S.~Kiranyaz, and M.~Gabbouj, ``Global ecg classification by
  self-operational neural networks with feature injection,'' \emph{arXiv
  preprint arXiv:2204.03768}, 2022.

\bibitem{ince2022blind}
T.~Ince, S.~Kiranyaz, O.~C. Devecioglu, M.~S. Khan, M.~Chowdhury, and
  M.~Gabbouj, ``Blind restoration of real-world audio by 1d operational gans,''
  \emph{arXiv preprint arXiv:2212.14618}, 2022.

\bibitem{malik2021bm3d}
J.~Malik, S.~Kiranyaz, M.~Yamac, and M.~Gabbouj, ``Bm3d vs 2-layer onn,'' in
  \emph{2021 IEEE International Conference on Image Processing (ICIP)}.\hskip
  1em plus 0.5em minus 0.4em\relax IEEE, 2021, pp. 1994--1998.

\bibitem{ulyanov2016instance}
D.~Ulyanov, A.~Vedaldi, and V.~Lempitsky, ``Instance normalization: The missing
  ingredient for fast stylization,'' \emph{arXiv preprint arXiv:1607.08022},
  2016.

\bibitem{ioffe2015batch}
S.~Ioffe and C.~Szegedy, ``Batch normalization: Accelerating deep network
  training by reducing internal covariate shift,'' in \emph{International
  conference on machine learning}.\hskip 1em plus 0.5em minus 0.4em\relax PMLR,
  2015, pp. 448--456.

\bibitem{wang2020frequency}
L.~Wang, T.~Bi, and Y.~Shi, ``A frequency-separated 3d-cnn for hyperspectral
  image super-resolution,'' \emph{IEEE Access}, vol.~8, pp. 86\,367--86\,379,
  2020.

\bibitem{kingma2014adam}
D.~P. Kingma and J.~Ba, ``Adam: A method for stochastic optimization,''
  \emph{arXiv preprint arXiv:1412.6980}, 2014.

\bibitem{kiranyaz2017generalized}
S.~Kiranyaz, T.~Ince, A.~Iosifidis, and M.~Gabbouj, ``Generalized model of
  biological neural networks: progressive operational perceptrons,'' in
  \emph{2017 International Joint Conference on Neural Networks (IJCNN)}.\hskip
  1em plus 0.5em minus 0.4em\relax IEEE, 2017, pp. 2477--2485.

\bibitem{kiranyaz2017progressive}
------, ``Progressive operational perceptrons,'' \emph{Neurocomputing}, vol.
  224, pp. 142--154, 2017.

\bibitem{tran2020progressive}
D.~T. Tran, S.~Kiranyaz, M.~Gabbouj, and A.~Iosifidis, ``Progressive
  operational perceptrons with memory,'' \emph{Neurocomputing}, vol. 379, pp.
  172--181, 2020.

\bibitem{tran2019heterogeneous}
------, ``Heterogeneous multilayer generalized operational perceptron,''
  \emph{IEEE transactions on neural networks and learning systems}, vol.~31,
  no.~3, pp. 710--724, 2019.

\bibitem{tran2019knowledge}
------, ``Knowledge transfer for face verification using heterogeneous
  generalized operational perceptrons,'' in \emph{2019 IEEE International
  Conference on Image Processing (ICIP)}.\hskip 1em plus 0.5em minus
  0.4em\relax IEEE, 2019, pp. 1168--1172.

\bibitem{kiranyaz2021exploiting}
S.~Kiranyaz, J.~Malik, H.~B. Abdallah, T.~Ince, A.~Iosifidis, and M.~Gabbouj,
  ``Exploiting heterogeneity in operational neural networks by synaptic
  plasticity,'' \emph{Neural Computing and Applications}, vol.~33, pp.
  7997--8015, 2021.

\bibitem{malik2021self}
J.~Malik, S.~Kiranyaz, and M.~Gabbouj, ``Self-organized operational neural
  networks for severe image restoration problems,'' \emph{Neural Networks},
  vol. 135, pp. 201--211, 2021.

\bibitem{yilmaz2021self}
M.~A. Y{\'\i}lmaz, O.~Keles{\c{s}}, H.~G{\"u}ven, A.~M. Tekalp, J.~Malik, and
  S.~K{\'\i}ranyaz, ``Self-organized variational autoencoders (self-vae) for
  learned image compression,'' in \emph{2021 IEEE International Conference on
  Image Processing (ICIP)}.\hskip 1em plus 0.5em minus 0.4em\relax IEEE, 2021,
  pp. 3732--3736.

\bibitem{kelecs2021self}
O.~Kele{\c{s}}, A.~M. Tekalp, J.~Malik, and S.~K$\iota$ranyaz, ``Self-organized
  residual blocks for image super-resolution,'' in \emph{2021 IEEE
  International Conference on Image Processing (ICIP)}.\hskip 1em plus 0.5em
  minus 0.4em\relax IEEE, 2021, pp. 589--593.

\bibitem{soltanian2021speech}
M.~Soltanian, J.~Malik, J.~Raitoharju, A.~Iosifidis, S.~Kiranyaz, and
  M.~Gabbouj, ``Speech command recognition in computationally constrained
  environments with a quadratic self-organized operational layer,'' in
  \emph{2021 International Joint Conference on Neural Networks (IJCNN)}.\hskip
  1em plus 0.5em minus 0.4em\relax IEEE, 2021, pp. 1--6.

\bibitem{jiang2020generalized}
X.~Jiang, D.~Wang, D.~T. Tran, S.~Kiranyaz, M.~Gabbouj, and X.~Feng,
  ``Generalized operational classifiers for material identification,'' in
  \emph{2020 IEEE 22nd International Workshop on Multimedia Signal Processing
  (MMSP)}.\hskip 1em plus 0.5em minus 0.4em\relax IEEE, 2020, pp. 1--5.

\bibitem{kiranyaz2022blind}
S.~Kiranyaz, O.~C. Devecioglu, T.~Ince, J.~Malik, M.~Chowdhury, T.~Hamid,
  R.~Mazhar, A.~Khandakar, A.~Tahir, T.~Rahman \emph{et~al.}, ``Blind ecg
  restoration by operational cycle-gans,'' \emph{IEEE Transactions on
  Biomedical Engineering}, vol.~69, no.~12, pp. 3572--3581, 2022.

\bibitem{rahman2022robust}
A.~Rahman, M.~E. Chowdhury, A.~Khandakar, A.~M. Tahir, N.~Ibtehaz, M.~S.
  Hossain, S.~Kiranyaz, J.~Malik, H.~Monawwar, and M.~A. Kadir, ``Robust
  biometric system using session invariant multimodal eeg and keystroke
  dynamics by the ensemble of self-onns,'' \emph{Computers in Biology and
  Medicine}, vol. 142, p. 105238, 2022.

\end{thebibliography}

\begin{IEEEbiography}[{\includegraphics[width=1in,height=1.25in,clip,keepaspectratio]{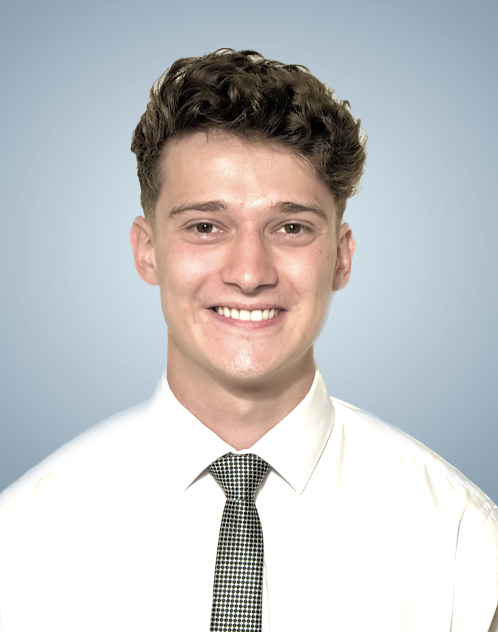}}]{Alexander Ulrichsen} received a first class MEng degree in Electronic and Electrical 
Engineering from the University of Strathclyde, Scotland, in 2020. He is currently pursuing a PhD 
degree in deep learning and image processing at the University of Strathclyde, Scotland.

His research interests include the 
development of hyperspectral image processing, deep learning, and camera-based tracking systems. 

He has published papers at precision livestock conferences as a PhD student and won the award for best student presentation at the Precision Dairy Conference 2022.
\end{IEEEbiography}

\begin{IEEEbiography}[{\includegraphics[width=1in,height=1.25in,clip,keepaspectratio]{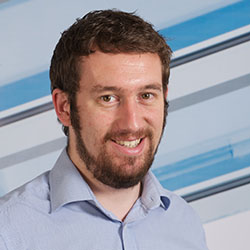}}]{Paul Murray} is a Senior Lecturer at the University of Strathclyde working in the area of image processing and hyperspectral imaging.

His research interests can be summarised as signal and image processing, hyperspectral imaging, machine learning, image stitching, object detection and tracking, mathematical morphology and the hit-or-miss-transform.
\end{IEEEbiography}

\begin{IEEEbiography}[{\includegraphics[width=1in,height=1.25in,clip,keepaspectratio]{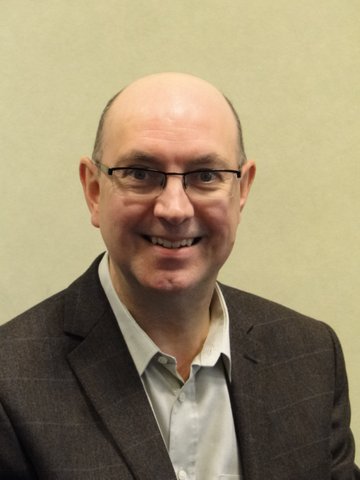}}]{Stephen Marshall} holds a first class honours degree in Electrical and Electronic Engineering from the University of Nottingham and a PhD in Image Processing from the University of Strathclyde. His recent research activities have focused on the application of novel signal and image processing techniques to Hyperspectral Imaging. He has published over 200 conference and journal papers on these topics including IEE, IEEE, SPIE, SIAM, ICASSP, VIE and EUSIPCO. He has also been a reviewer for these and other journals and conferences. He is a Fellow of the Institution of Engineering and Technology (IET). He has also been successful in obtaining research funding from National, International, and Industrial sources. These sources include EPSRC, EU, BBSRC, NERC and Innovate UK.
\end{IEEEbiography}

\begin{IEEEbiography}[{\includegraphics[width=1in,height=1.25in,clip,keepaspectratio]{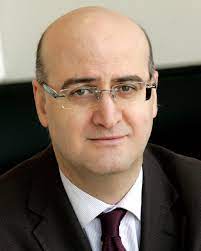}}]{Moncef Gabbouj} is a Professor of Signal Processing at the Department of Computing Sciences, Tampere University, Finland.

His research interests include Big Data analytics, multimedia content-based analysis, indexing and retrieval, artificial intelligence, machine learning, pattern recognition, nonlinear signal and image processing and analysis, voice conversion, and video processing and coding.
\end{IEEEbiography}

\begin{IEEEbiography}[{\includegraphics[width=1in,height=1.25in,clip,keepaspectratio]{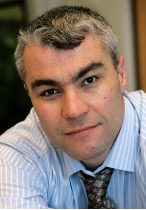}}]{Serkan Kiranyaz} was born in Turkey, 1972. He received his BS and MS degrees in the Electrical and Electronics Department at Bilkent University, Ankara, Turkey, in 1994 and 1996, respectively. He received his PhD degree in 2005 and his Docency in 2007 from Tampere University of Technology, Institute of Signal Processing respectively. He was working as a Professor in Signal Processing Department at the same university from 2009 to 2015. He currently works as a Professor at Qatar University, Doha, Qatar.

Prof. Kiranyaz has noteworthy expertise and background in various signal processing domains. He published two books, 7 book chapters, 10 patents/applications, more than 100 journal articles in several IEEE Transactions and other high-impact journals, and more than 120 papers in international conferences. He served as PI and LPI in several national and international projects. His principal research field is machine learning and signal processing. He is rigorously aiming for reinventing the ways in novel signal processing paradigms, enriching them with new approaches, especially in machine intelligence, and revolutionizing the means of “learn-to-process” signals. He made significant contributions to bio-signal analysis, particularly EEG and ECG analysis and processing, classification and segmentation, computer vision with applications to recognition, classification, multimedia retrieval, evolving systems and evolutionary machine learning, swarm intelligence and evolutionary optimization.
\end{IEEEbiography}

\begin{IEEEbiography}[{\includegraphics[width=1in,height=1.25in,clip,keepaspectratio]{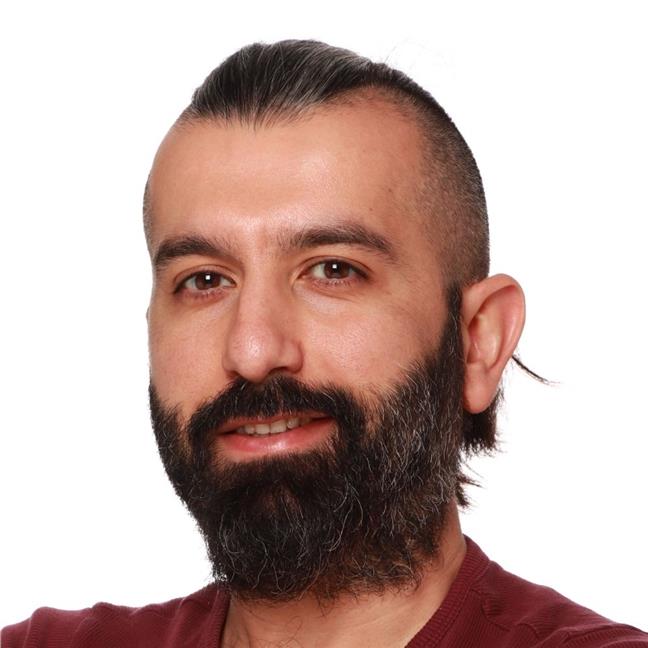}}]{Mehmet Yamaç} received a BS degree in electrical and electronics engineering from Anadolu University, Eski¸sehir, Turkey, in 2009, and a MS degree in electrical and electronics engineering from Bogaziçi University, Istanbul, Turkey, in 2014. He is currently pursuing a PhD degree with the Department of Computing Sciences, Tampere University, Tampere, Finland. He was a Research and Teaching Assistant with Bogazici University from 2012 to 2017 and a Researcher with Tampere University from 2017 to 2020. He is currently a Senior Researcher with Huawei Technologies Oy, Tampere. He has co-authored more than 30 papers, two of them nominated for the “Best (or Student Best) Paper Award” in EUVIP 2018 and EUSIPCO 2019. His research interests are computer and machine vision, machine learning, and compressive sensing.
\end{IEEEbiography}

\begin{IEEEbiography}[{\includegraphics[width=1in,height=1.25in,clip,keepaspectratio]{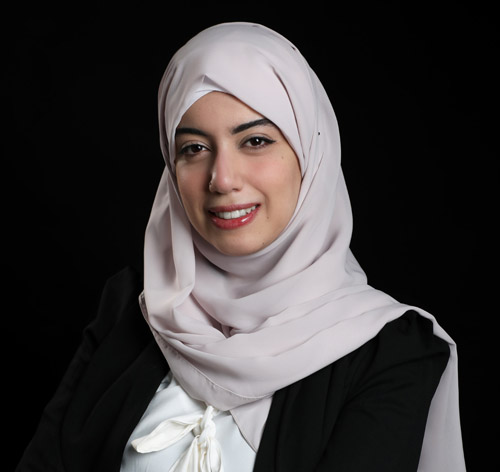}}]{Nour Aburaed} received a BS degree in Computer Engineering from Khalifa University of Science and Technology, Abu Dhabi, UAE, in 2014 and a MS degree in Electrical and Computer Engineering from the same university in 2016, particularly specialized in High-ISO image de-noising and Quantum Image Processing.
From 2016 to 2018, she was a Teaching Assistant at Khalifa University of Science and Technology for various theoretical and laboratory-based courses, including Calculus and Physics. She has been a Research Assistant at the Mohammed Bin Rashid Space Centre (MBRSC) Laboratory based at the University of Dubai since June 2018, where she applies image processing and artificial intelligence techniques within the context of remote sensing. Nour is also currently a PhD student at the University of Strathclyde, where she researches spatial enhancement techniques of hyperspectral remote sensing imagery. Her interests include Hyperspectral Imagery, Super Resolution, Object Detection, Semantic Segmentation, and Convolutional Neural Networks. Nour was the recipient of the President’s Scholarship and Master Research Teaching Scholarship (MRTS) from Khalifa University of Science and Technology for International Students.
\end{IEEEbiography}

\clearpage
\appendix

\section*{Self-ONNs}\label{appendix:Self-ONN}

Generalized Operational Perceptrons (GOPs) \cite{kiranyaz2017generalized, kiranyaz2017progressive, tran2020progressive, tran2019heterogeneous, tran2019knowledge} have recently been developed towards the goal of modeling biological neurons with \textit{distinct} synaptic connections. GOPs have demonstrated a superior diversity, as encountered in biological neural networks, which resulted in an elegant performance level on numerous challenging problems where conventional MLPs entirely failed. Following in the GOP's footsteps, Operational Neural Networks (ONNs) \cite{kiranyaz2020operational, kiranyaz2021exploiting, malik2020fastonn} were developed as a superset of CNNs. ONNs not only outperform CNNs significantly, but they are also able to learn certain problems where CNNs fail entirely. However, ONNs also exhibit certain drawbacks such as strict dependability to the operators in the operator set library, the mandatory search for the best operator set for each layer/neuron, and the need for setting (fixing) the operator sets of the output layer neuron(s) in advance. The operator diversity is also limited since a single operator set is assigned one or usually more neurons and this makes all (synaptic) connections have the same operator.  

\begin{figure*}[!b]
    \centering
    \includegraphics[width=1\textwidth]{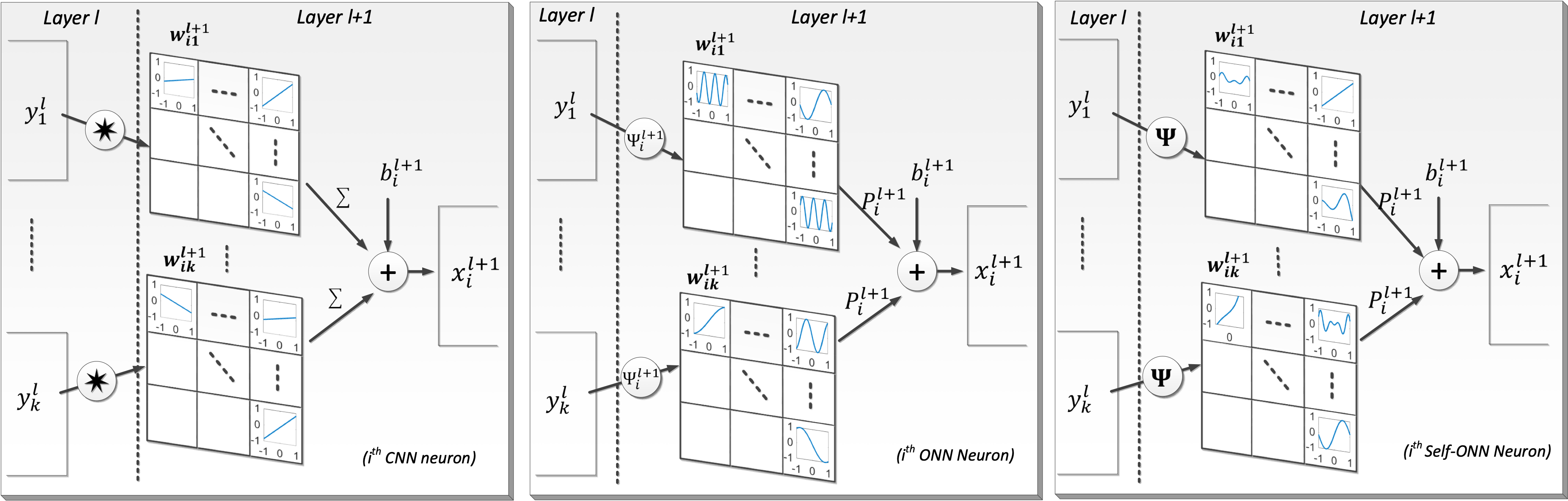}
    \caption{An illustration from [9] of the nodal operations in the kernels of the ith CNN (left), ONN (middle), and Self-ONN (right) neurons at layer $l+1$.}
    \label{fig:Self-ONN}
\end{figure*}

Furthermore, the operator set for the “right” transformation may or may not exist in the library. For this purpose, “Self-Organizing” ONNs (Self-ONNs) \cite{kiranyaz2021self} were recently proposed with the \textit{generative} neuron model that addresses this drawback by customizing each nodal operator on the fly. This is in fact the case for biological neurons where the synaptic connections can exhibit any arbitrary form or pattern. In brief, a generative neuron is basically an operational neuron with a \textit{composite} nodal operator that can be generated during training without any restrictions. As a result, with such generative neurons, a Self-ONN can self-organize its nodal operators during training, and thus, it will have the nodal operator functions “optimized” by the training process to maximize the learning performance. For instance, in the sample illustration shown in Figure \ref{fig:Self-ONN}, the CNN and ONN neurons have \textit{static} nodal operators (linear and harmonic, respectively) for their 3x3 kernels, while the generative neuron can have \textit{any} arbitrary nodal function, $\boldsymbol{\Psi}$,  (including possibly standard types such as linear and harmonic functions) for each kernel element of each connection. This is a great flexibility that permits the formation of \textit{any} nodal operator function and also allows the creation of the optimal nodal operators during training to maximize the learning performance. 
As illustrated in Figure \ref{fig:Self-ONN} (middle), for conventional ONNs the input map of the ith neuron at the layer $l+1, x_i^(l+1)$, is composed in Eq. (\ref{eq:ONN}): 

\begin{equation}
\label{eq:ONN}
\begin{aligned}
    	& x_{i}^{l+1}=b_{i}^{l+1}+\sum_{k=1}^{N_{l}} \left( \textbf{oper2D}(y_{k}^{l},w_{ik}^{l+1} ,^{'} NoZeroPad^{'} \right) \\
    	& x^{l+1}_i(m, n) |_{(0,0)}^{(M-1,N-1)} = b^{l+1}_i + \\
	& \;\;\; \sum_{i=1}^{N_{l-1}} \left( P^{l+1}_i \left[ \begin{aligned} & \Psi^{l+1}_i \left( y^l_k(m, n), w^{l+1}_{ik}(0,0) \right),\, \dots\, ,  \\ & \Psi^{l+1}_i \left( y^l_k(m + r, n + t), w^{l+1}_{ik}(r,t) \right) ,\, \dots \end{aligned} \right] \right)
\end{aligned}
\end{equation}

where $y_{k}^{l}$  are the final output maps of the previous layer neurons \textit{operated} with the corresponding kernels, $w_{ik}^{l+1}$, with a particular nodal function, $\Psi_{i}^{l+1}$ such as linear (multiplication), sinusoid, exponential, Gaussian, chirp, Hermitian, etc. A close look at Eq. (\ref{eq:ONN}) reveals the fact that when the pool operator is “summation”, $P_{i}^{l+1} = \Sigma$, and the nodal operator is “linear”, $\Psi_{i}^{l+1} \left(y_{k}^{l} (m+r,n+t),w_{ik}^{l+1} (r,t) \right) = y_k^l (m+r,n+t) \times w_{ik}^{l+1} (r,t)$, for \textit{all} neurons, then the resulting homogenous ONN will be identical to a CNN. Hence, ONNs are indeed a superset of CNNs as the GOPs are a superset of MLPs. 

Self-ONNs with generative neurons differ from ONNs by the following two points: 
\begin{enumerate}
    \item Each “fixed-in-advance” nodal operator function with a scalar kernel element, $\Psi_{i}^{l+1} (y_k^l (m+r,n+t),w_{ik}^{l+1} (r,t))$, is \textit{approximated} by the composite nodal operator, $\boldsymbol{\Psi}(y_{k}^{l} (m+r,n+t),\boldsymbol{w_{ik}^{l+1}} (r,t))$, as expressed by the Maclaurin series,
    \item The scalar kernel parameter, $w_{ik}^{l+1} (r,t)$, of the kernel of an ONN neuron, is replaced by a Q-dimensional array, $\boldsymbol{w_{ik}^{l+1}} (r,t)$.
\end{enumerate}

In this way, any nodal operator function can be approximated with Maclaurin series near the origin as shown in Eq. (\ref{eq:Self-ONN}): 

\begin{equation}
\label{eq:Self-ONN}
    \boldsymbol{\Psi}(y,\boldsymbol{w})=w_{0} + w_{1}y + w_{2}y^{2}+\dots+w_{Q}y^{Q}
\end{equation}

where $w_q=\frac{f^{(q)} (0)}{q!}$ is the $q^{th}$ coefficient of the $Q^{th}$ order polynomial. During the back-propagation (BP) training, each $w_{q}$ of a kernel element is optimized for the learning problem at hand. Thanks to this ability, there is no need for any operator search for Self-ONNs and arbitrary nodal operators can be customized by the training process as illustrated in Figure \ref{fig:Self-ONN} (right). This results in enhanced flexibility and diversity over an operational neuron where only a standard nodal operator function has to be used for all kernels, each connected to an output map of a neuron in the previous layer. With this ability, in various 1D and 2D applications, Self-ONNs outperformed both conventional ONNs and CNNs with a significant gap \cite{kiranyaz2021self, malik2021self, devecioglu2021real, malik2021real, kiranyaz2021robust, yilmaz2021self, kelecs2021self, soltanian2021speech, jiang2020generalized, kiranyaz2022blind, rahman2022robust}.

\section*{Training Plots}\label{appendix:trainPlots}
\subsection{Pavia University}

% Subfigure display
\begin{figure}[H]
    \centering
    \subfloat[Training loss plot.]{
        \includegraphics[width=0.5\textwidth]{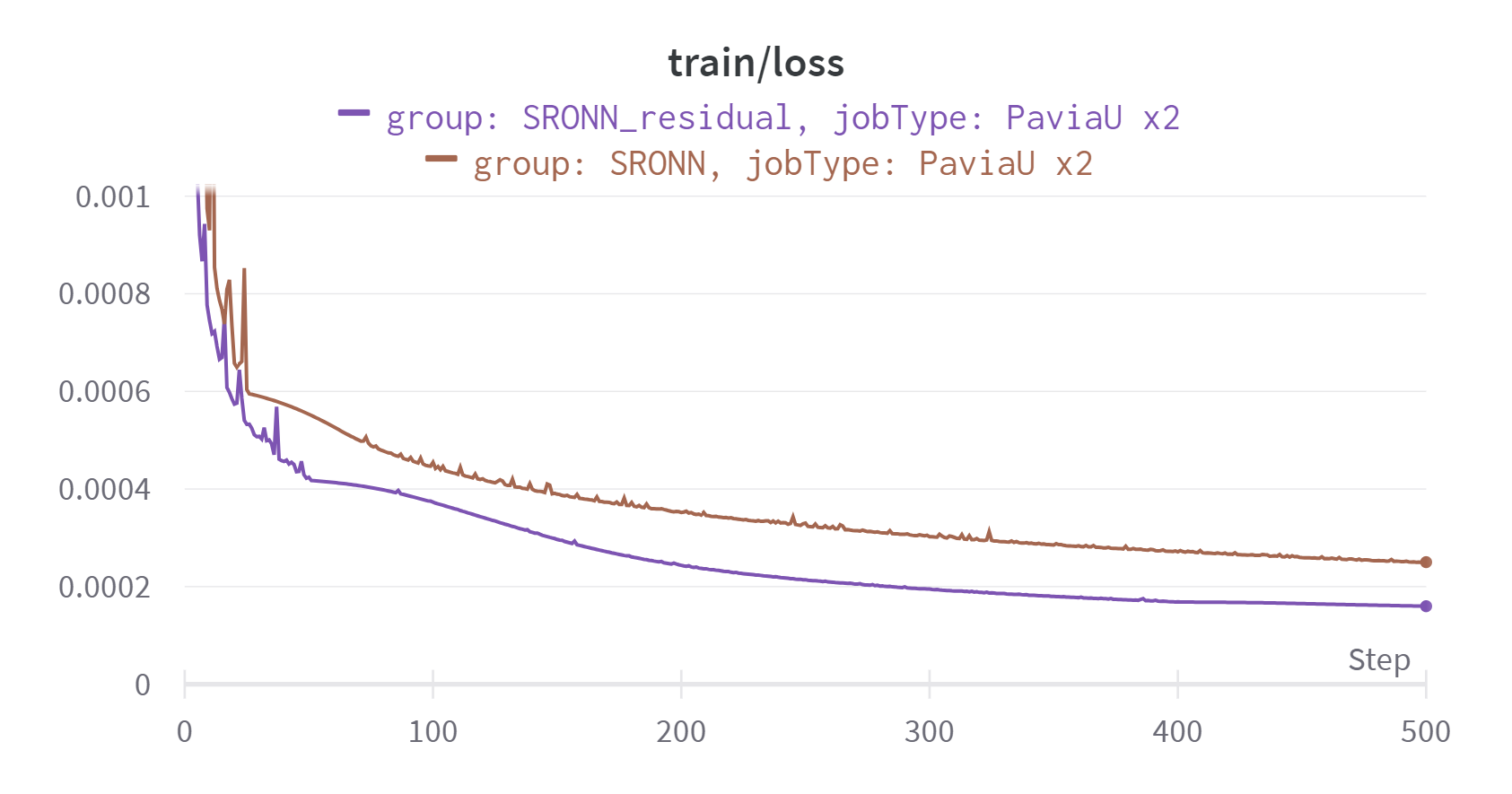}
        \label{fig:PaviaU_loss}
    }
    
    \subfloat[Validation SSIM plot.]{
        \includegraphics[width=0.5\textwidth]{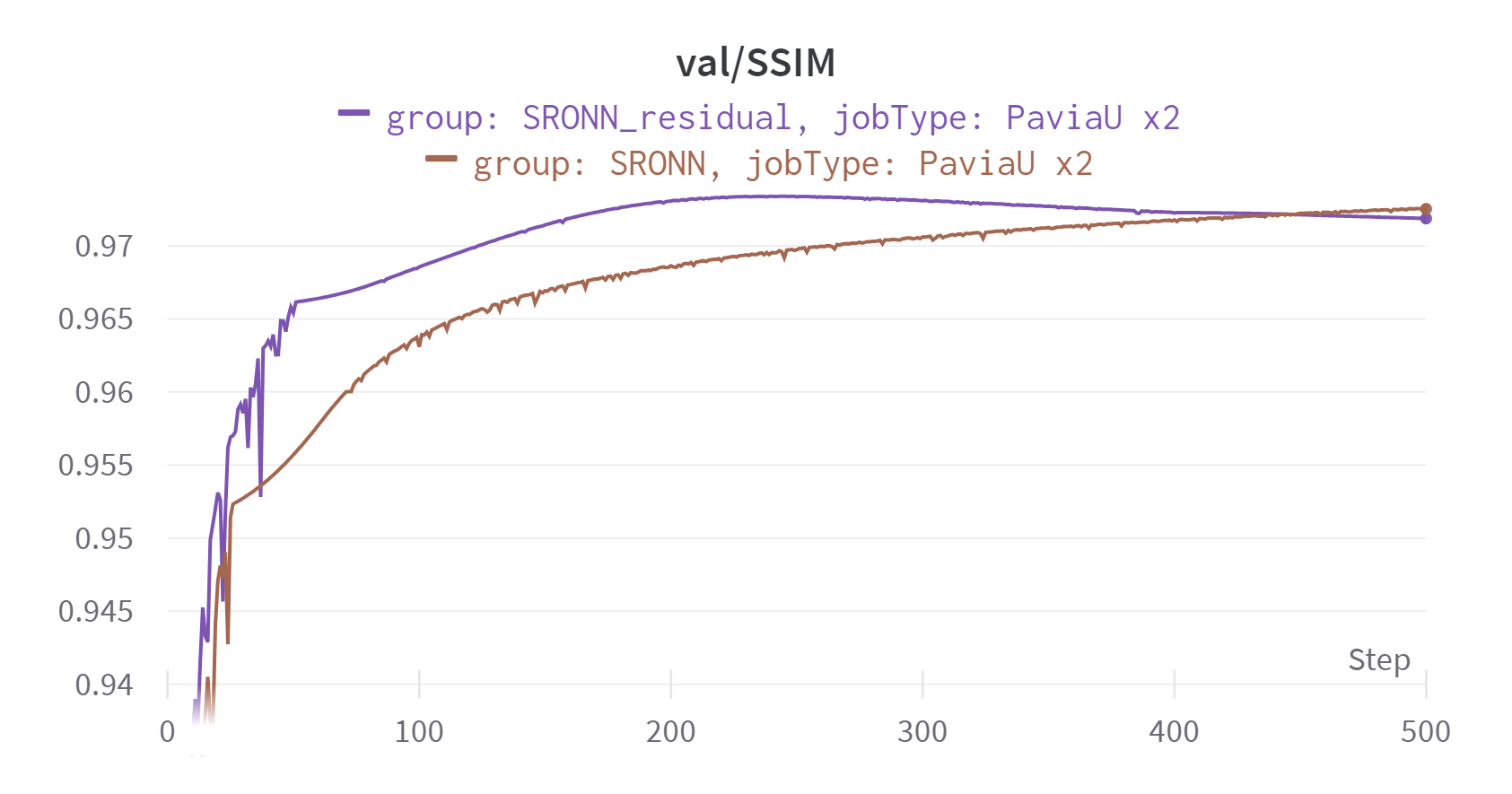}
        \label{fig:PaviaU_ssim}
    }
    
    \subfloat[Validation PSNR plot.]{
        \includegraphics[width=0.5\textwidth]{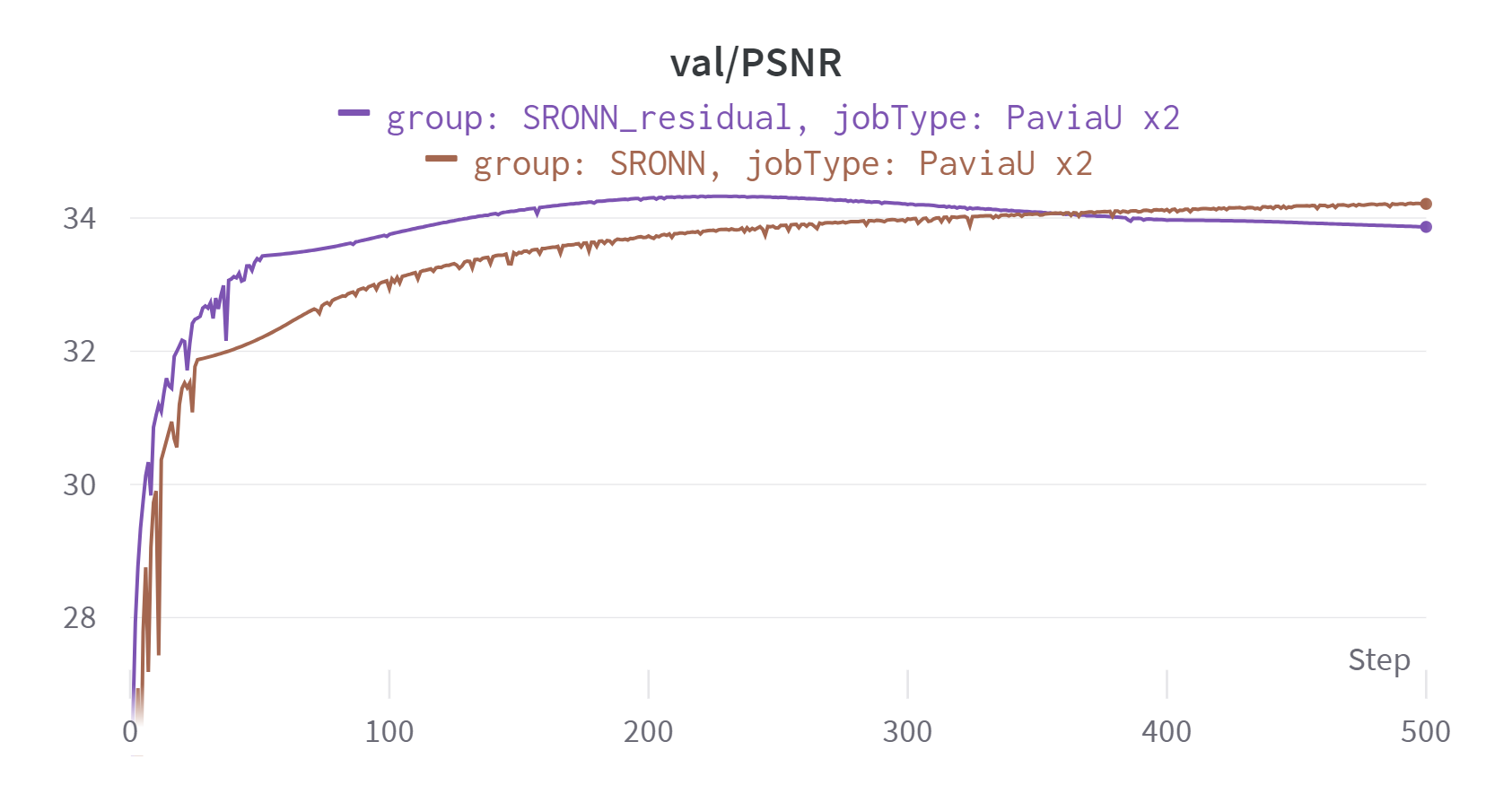}
        \label{fig:PaviaU_psnr}
    }
    
    \subfloat[Validation SAM plot.]{
        \includegraphics[width=0.5\textwidth]{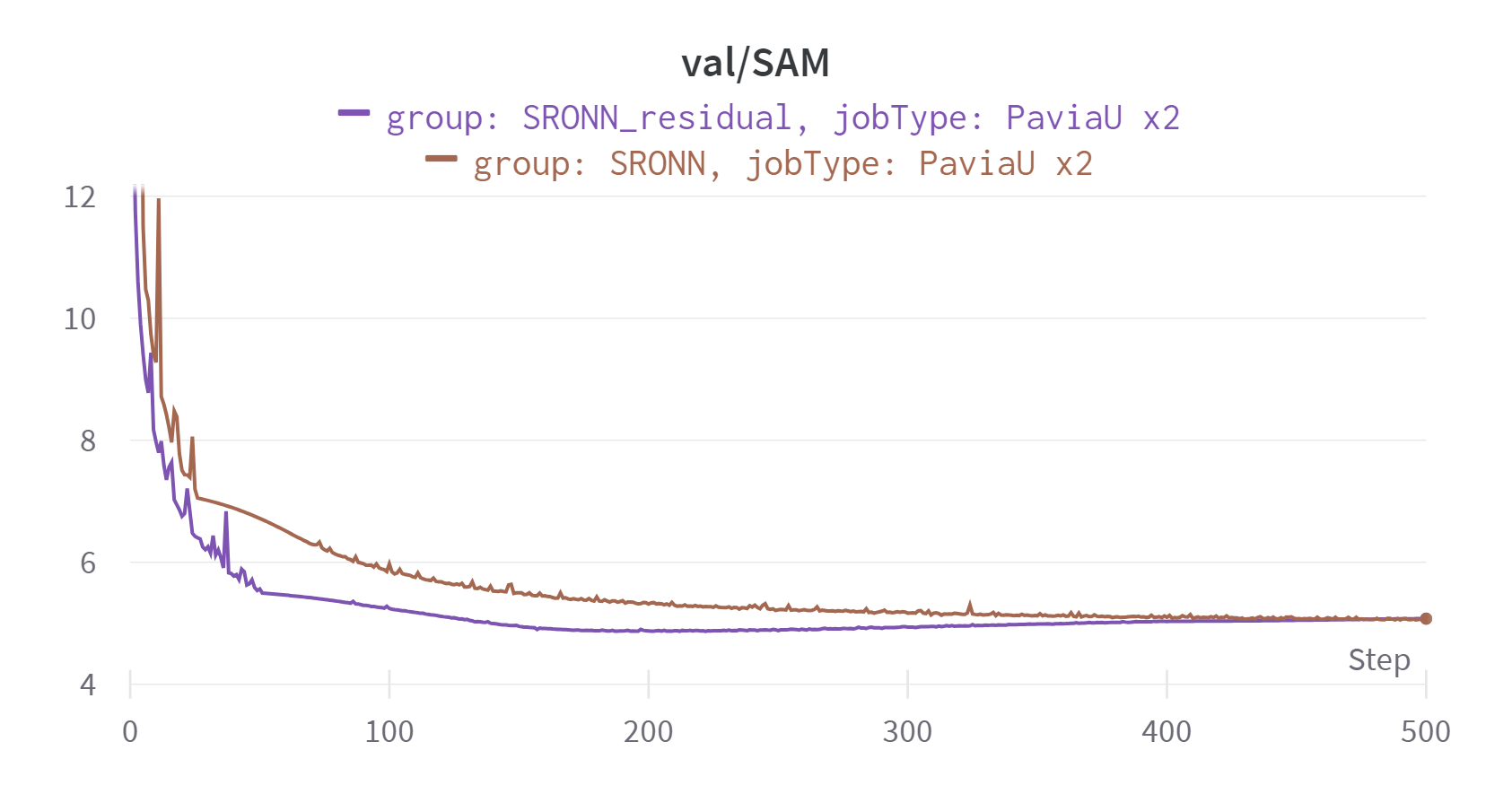}
        \label{fig:PaviaU_sam}
    }
    
    \caption{SRONN training and validation plots on the Pavia University dataset with and without a residual connection.}
    \label{fig:PaviaU_plots}
\end{figure}

\subsection{Salinas}

\begin{figure}[H]
    \centering
    \subfloat[Training loss plot.]{
        \includegraphics[width=0.5\textwidth]{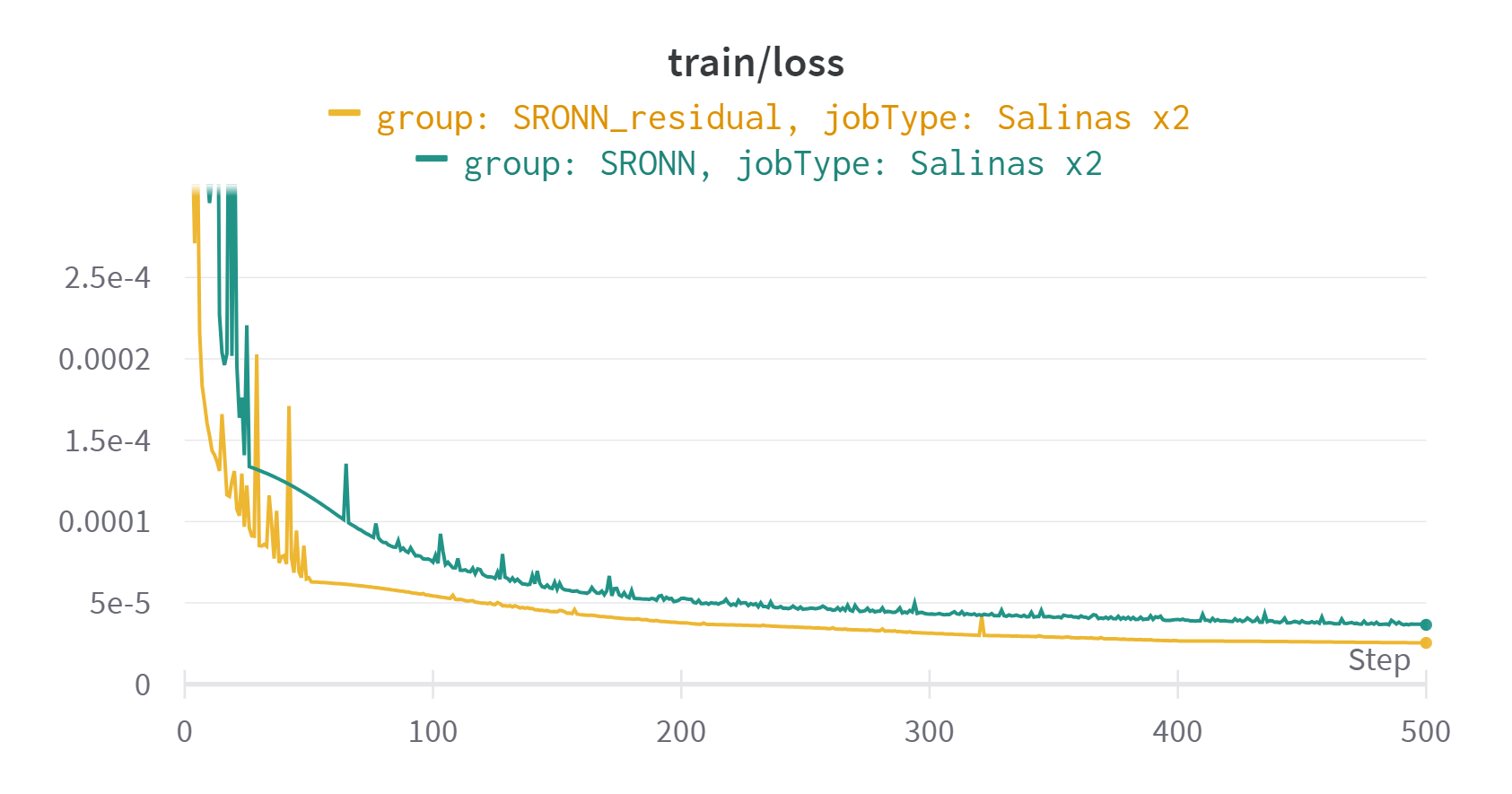}
        \label{fig:Salinas_loss}
    }
    
    \subfloat[Validation SSIM plot.]{
        \includegraphics[width=0.5\textwidth]{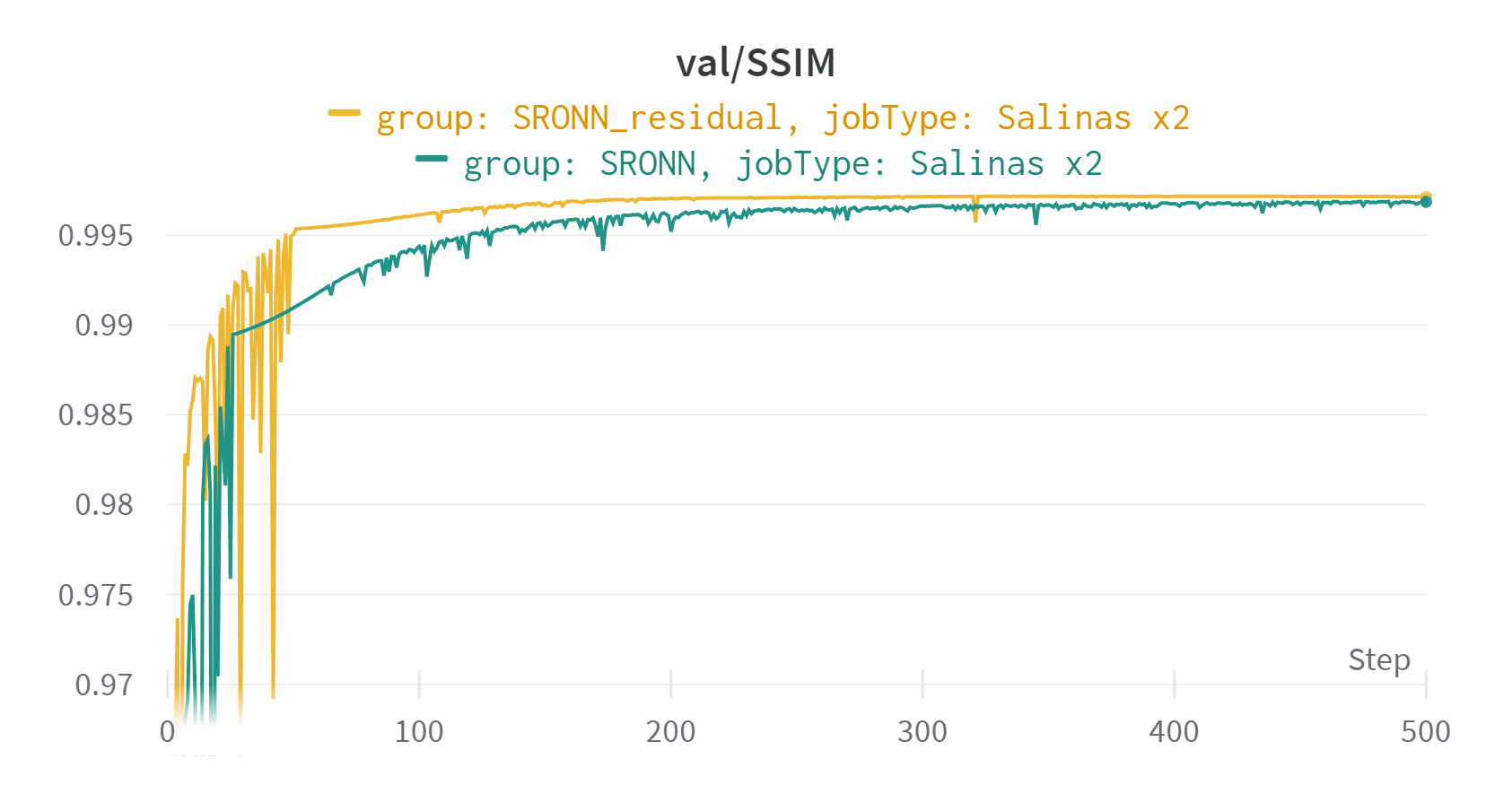}
        \label{fig:Salinas_ssim}
    }
    
    \subfloat[Validation PSNR plot.]{
        \includegraphics[width=0.5\textwidth]{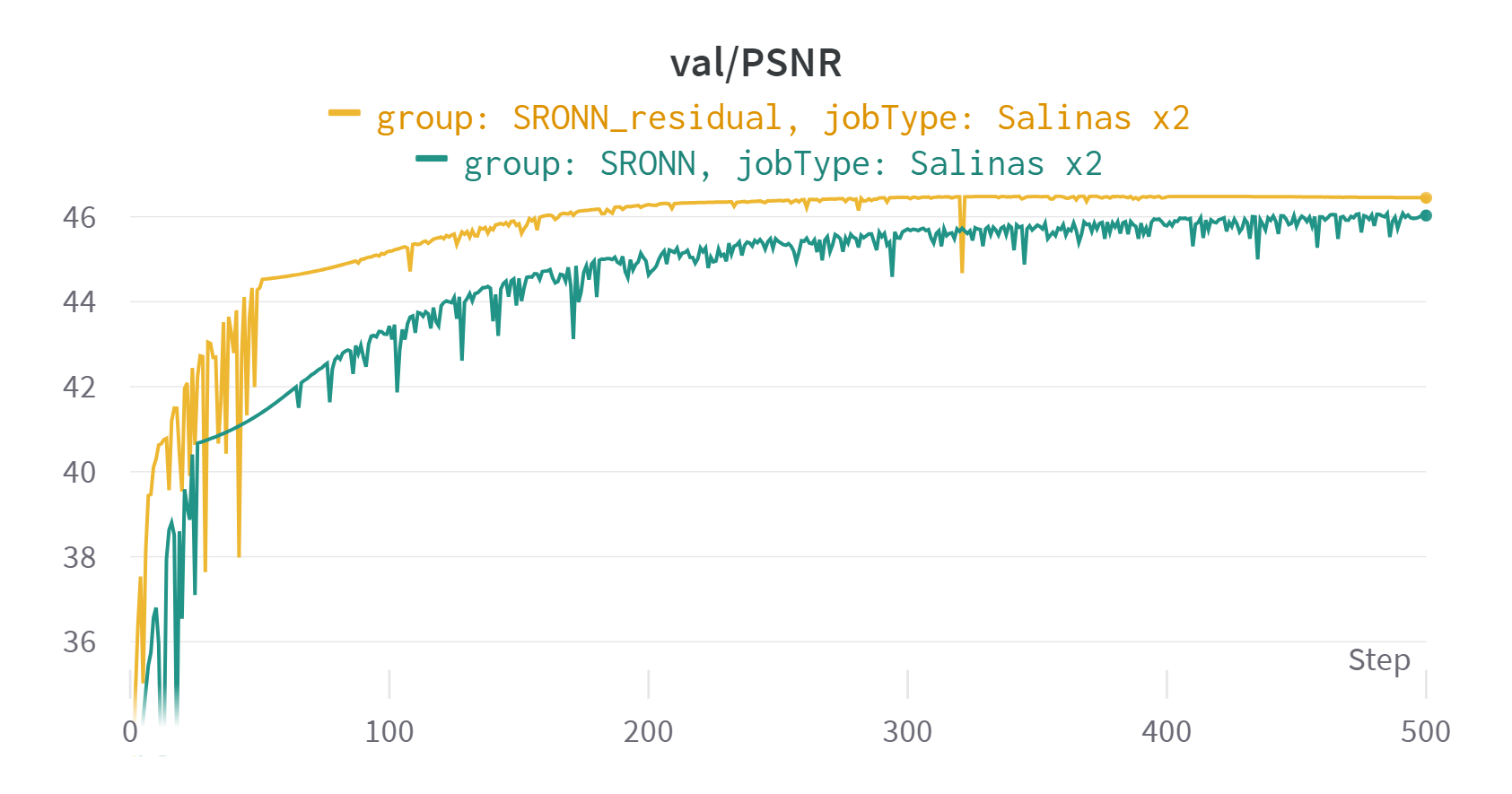}
        \label{fig:Salinas_psnr}
    }
    
    \subfloat[Validation SAM plot.]{
        \includegraphics[width=0.5\textwidth]{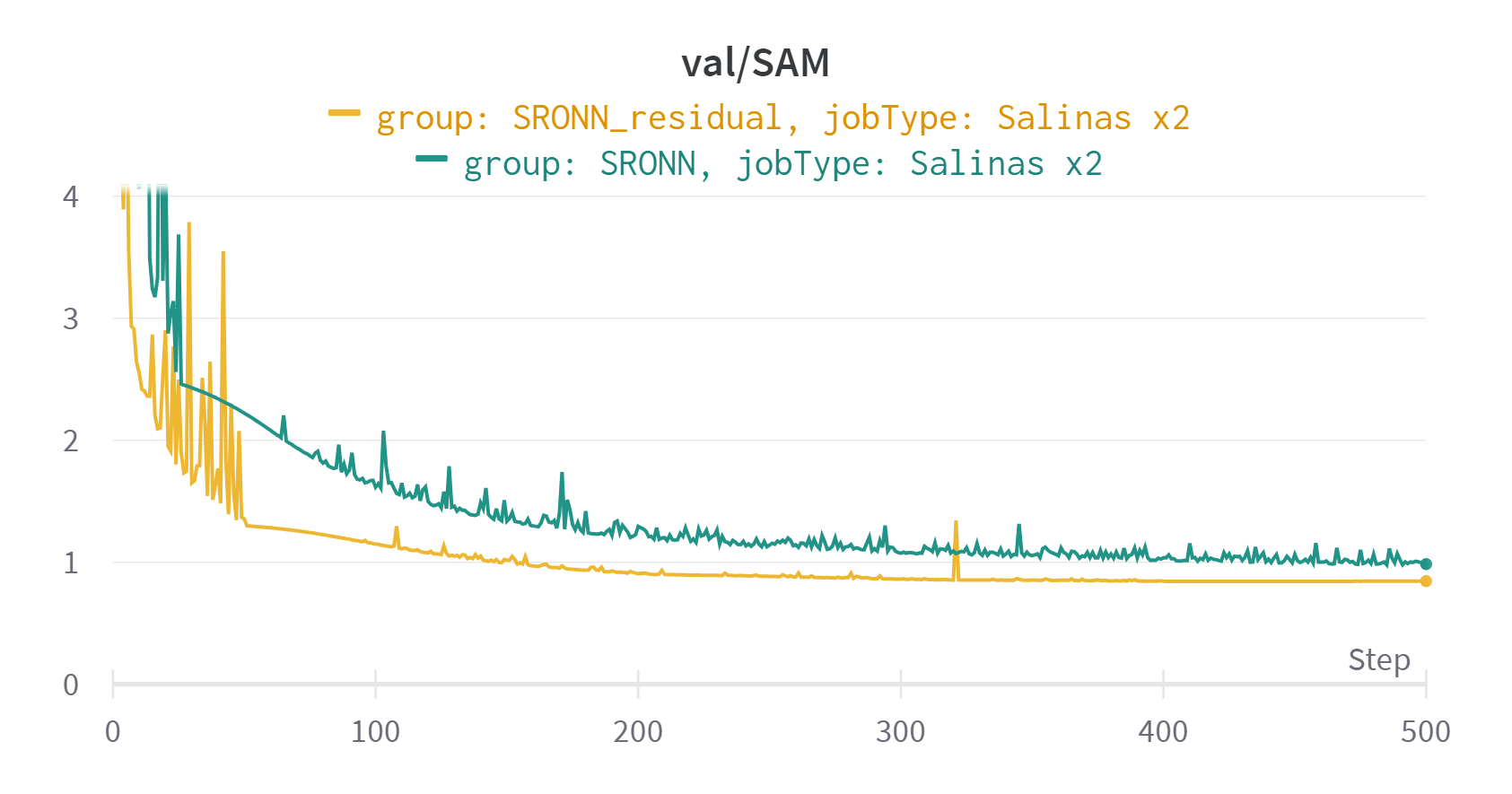}
        \label{fig:Salinas_sam}
    }
    
    \caption{SRONN training and validation plots on the Salinas dataset with and without a residual connection.}
    \label{fig:Salinas_plots}
\end{figure}

\subsection{Cuprite}
\begin{figure}[H]
    \centering
    \subfloat[Training loss plot.]{
        \includegraphics[width=0.5\textwidth]{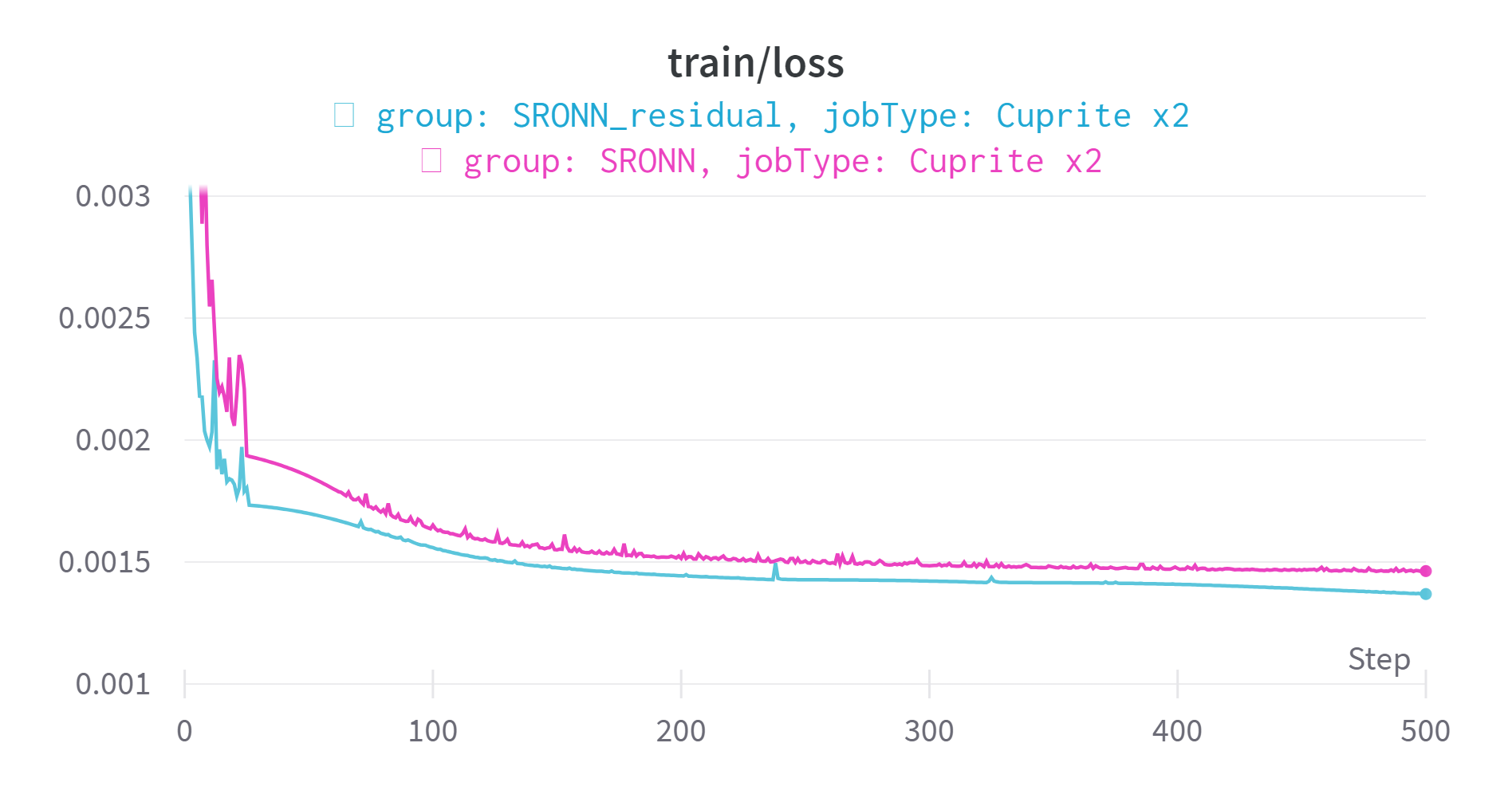}
        \label{fig:Cuprite_loss}
    }
    
    \subfloat[Validation SSIM plot.]{
        \includegraphics[width=0.5\textwidth]{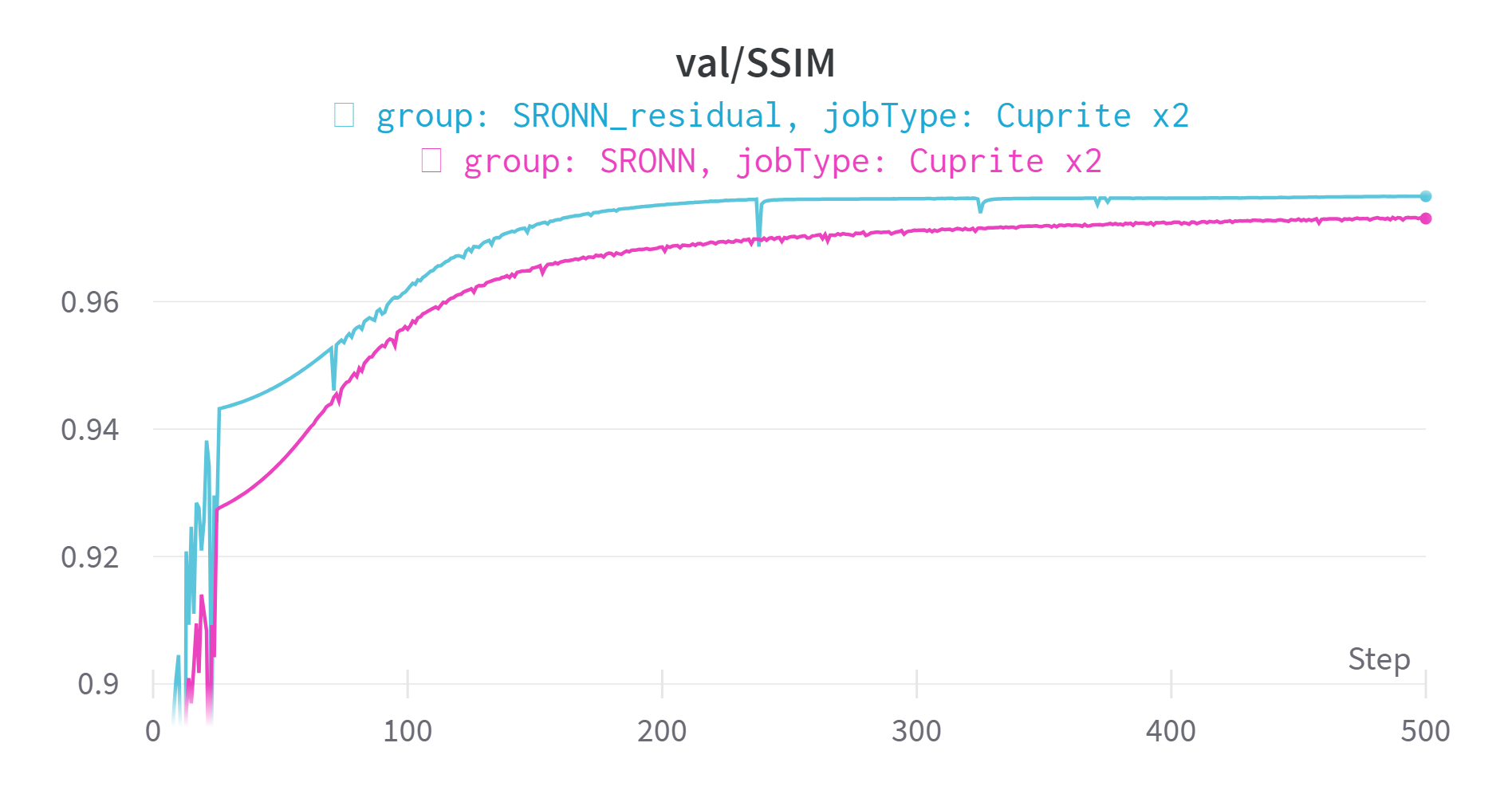}
        \label{fig:Cuprite_ssim}
    }
    
    \subfloat[Validation PSNR plot.]{
        \includegraphics[width=0.5\textwidth]{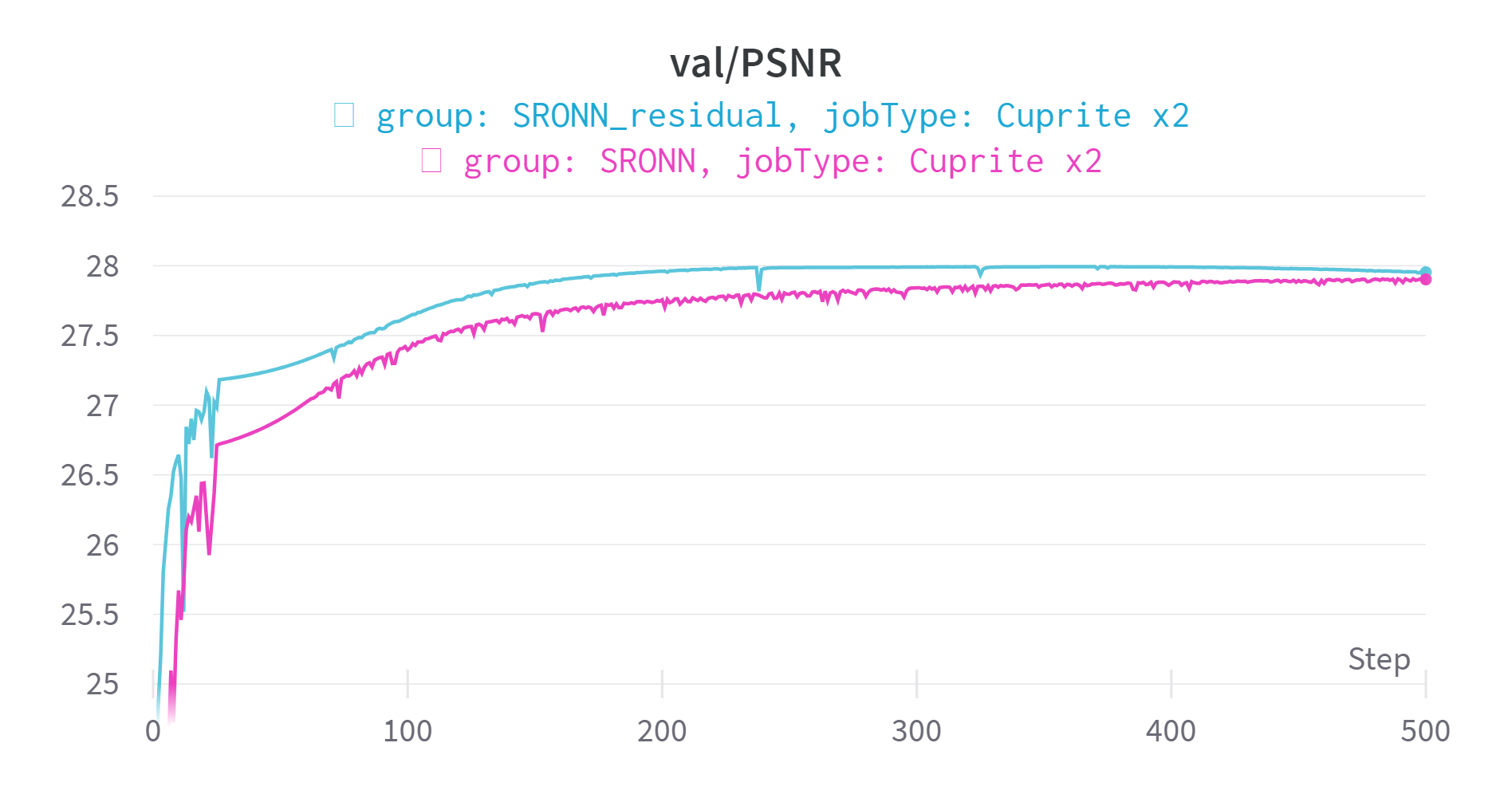}
        \label{fig:Cuprite_psnr}
    }
    
    \subfloat[Validation SAM plot.]{
        \includegraphics[width=0.5\textwidth]{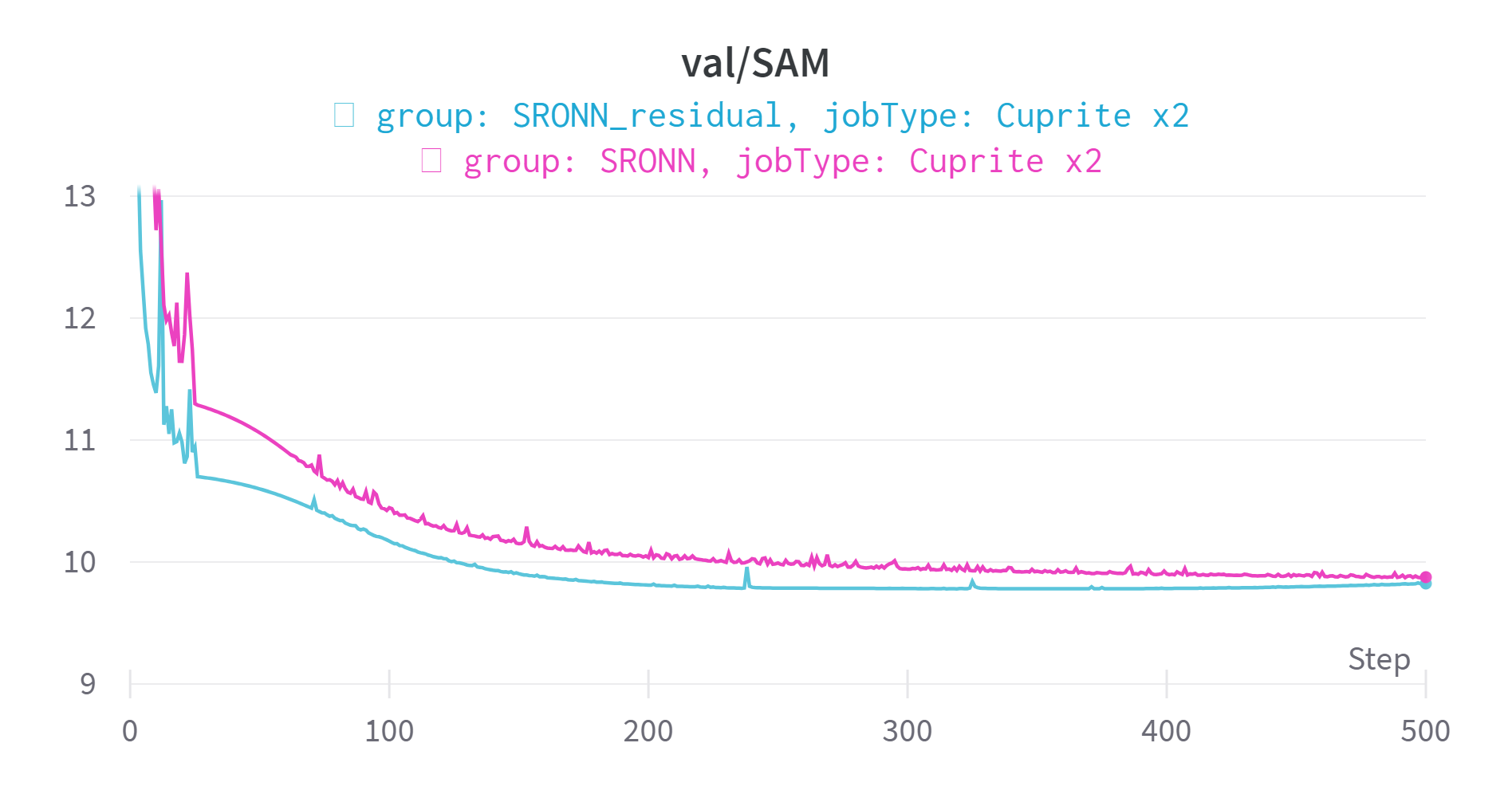}
        \label{fig:Cuprite_sam}
    }
    
    \caption{SRONN training and validation plots on the Cuprite dataset with and without a residual connection.}
    \label{fig:Cuprite_plots}
\end{figure}

\subsection{Urban}

\begin{figure}[H]
    \centering
    \subfloat[Training loss plot.]{
        \includegraphics[width=0.5\textwidth]{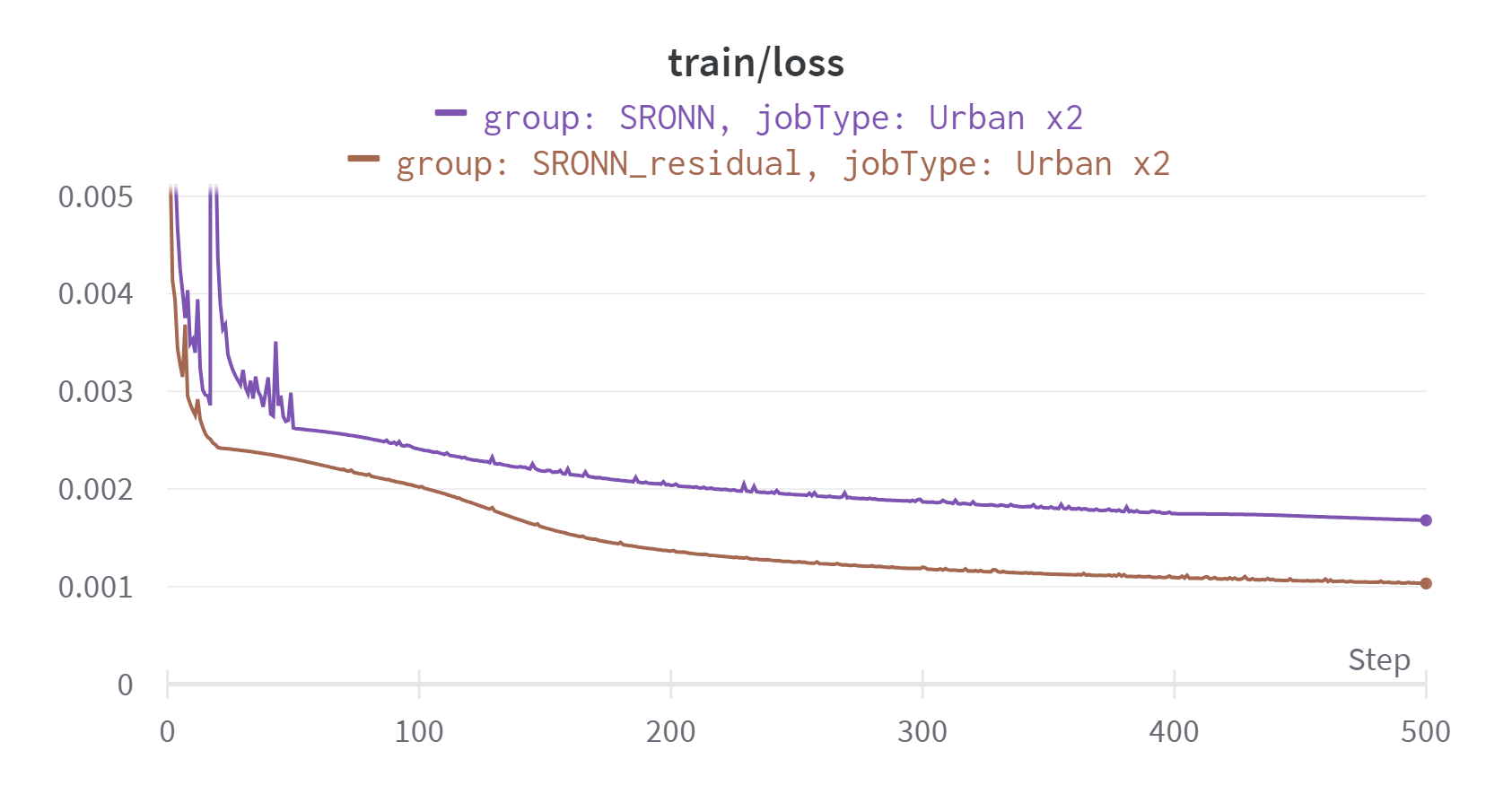}
        \label{fig:Urban_loss}
    }
    
    \subfloat[Validation SSIM plot.]{
        \includegraphics[width=0.5\textwidth]{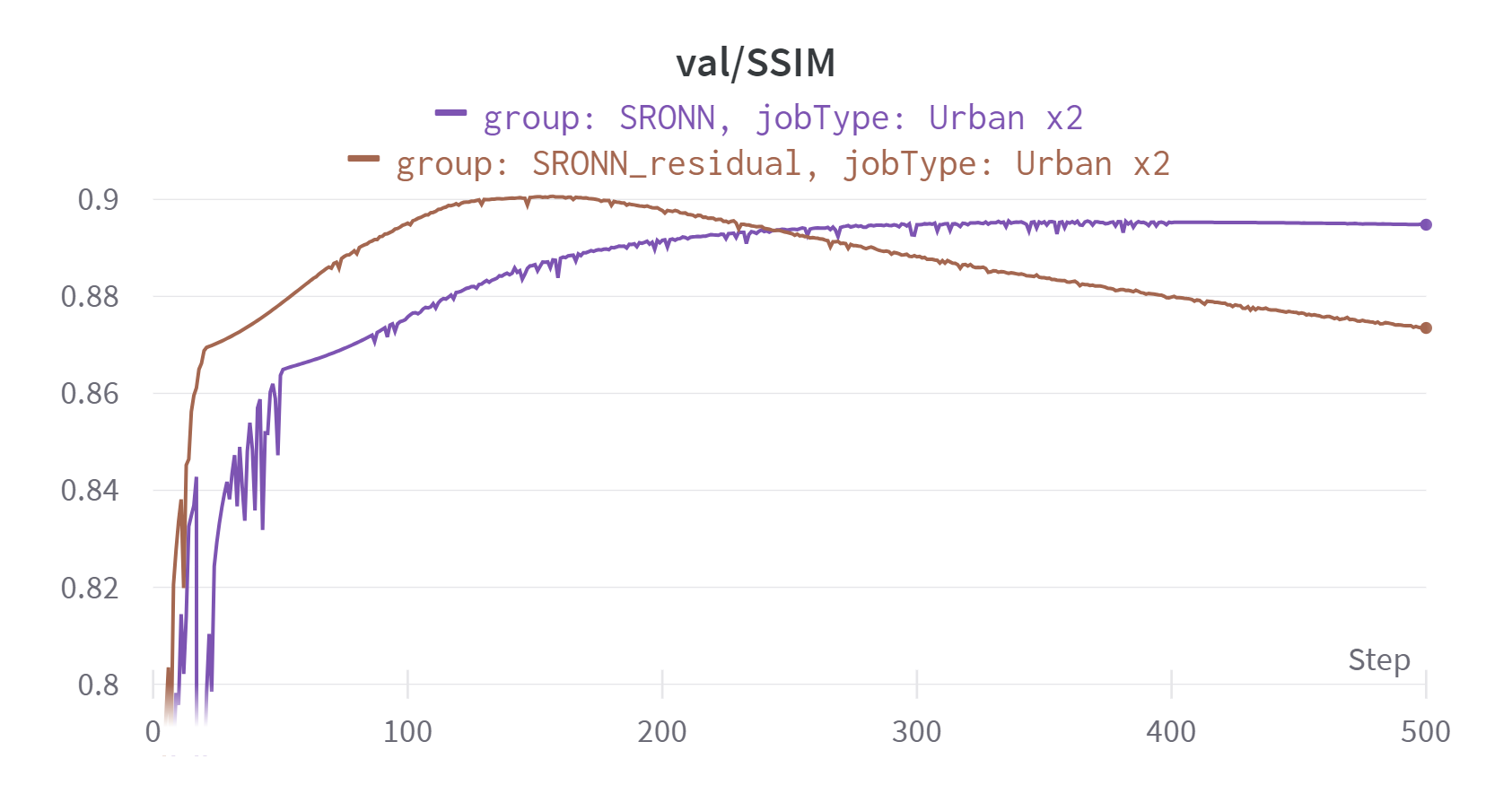}
        \label{fig:Urban_ssim}
    }
    
    \subfloat[Validation PSNR plot.]{
        \includegraphics[width=0.5\textwidth]{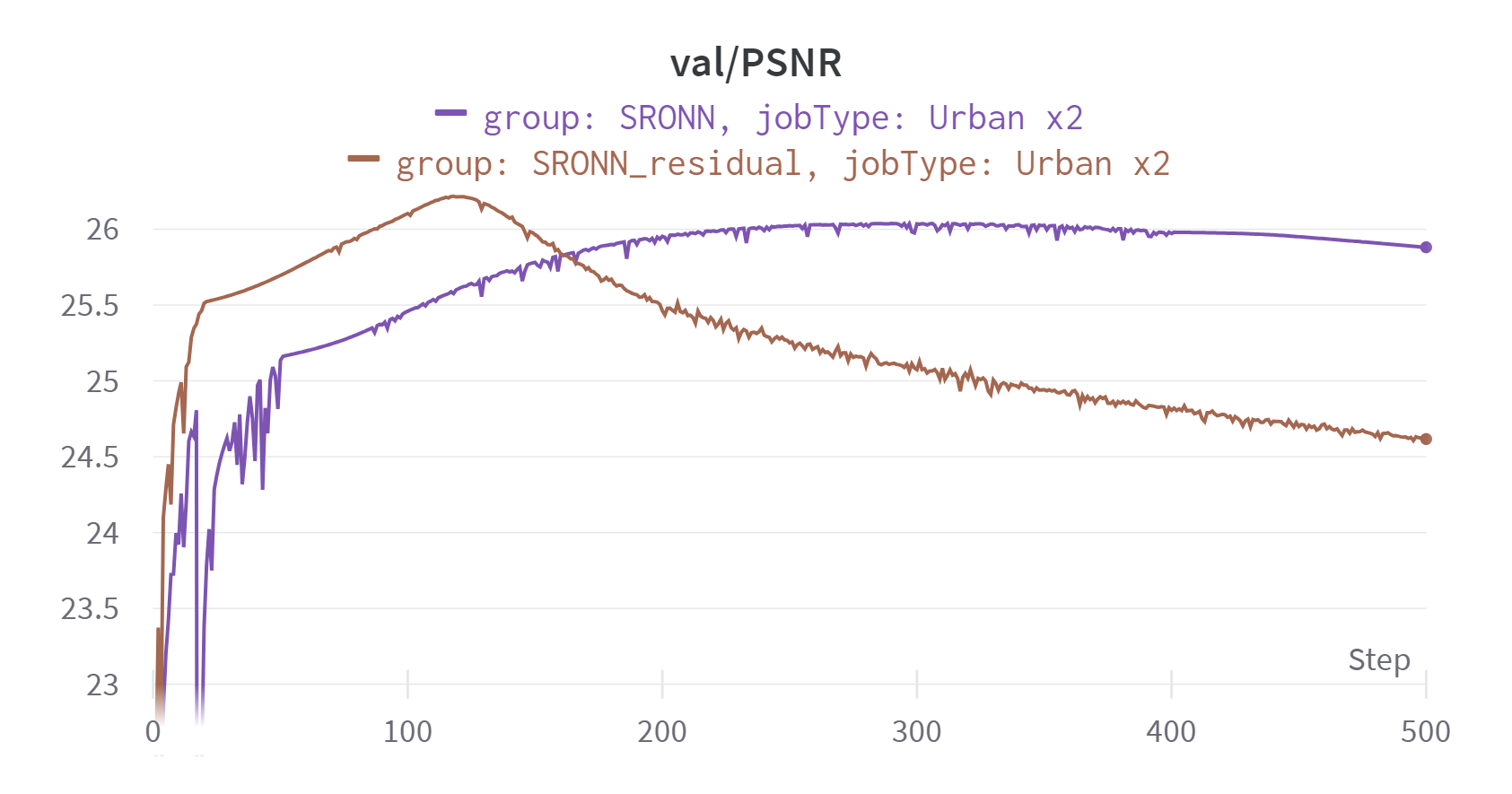}
        \label{fig:Urban_psnr}
    }
    
    \subfloat[Validation SAM plot.]{
        \includegraphics[width=0.5\textwidth]{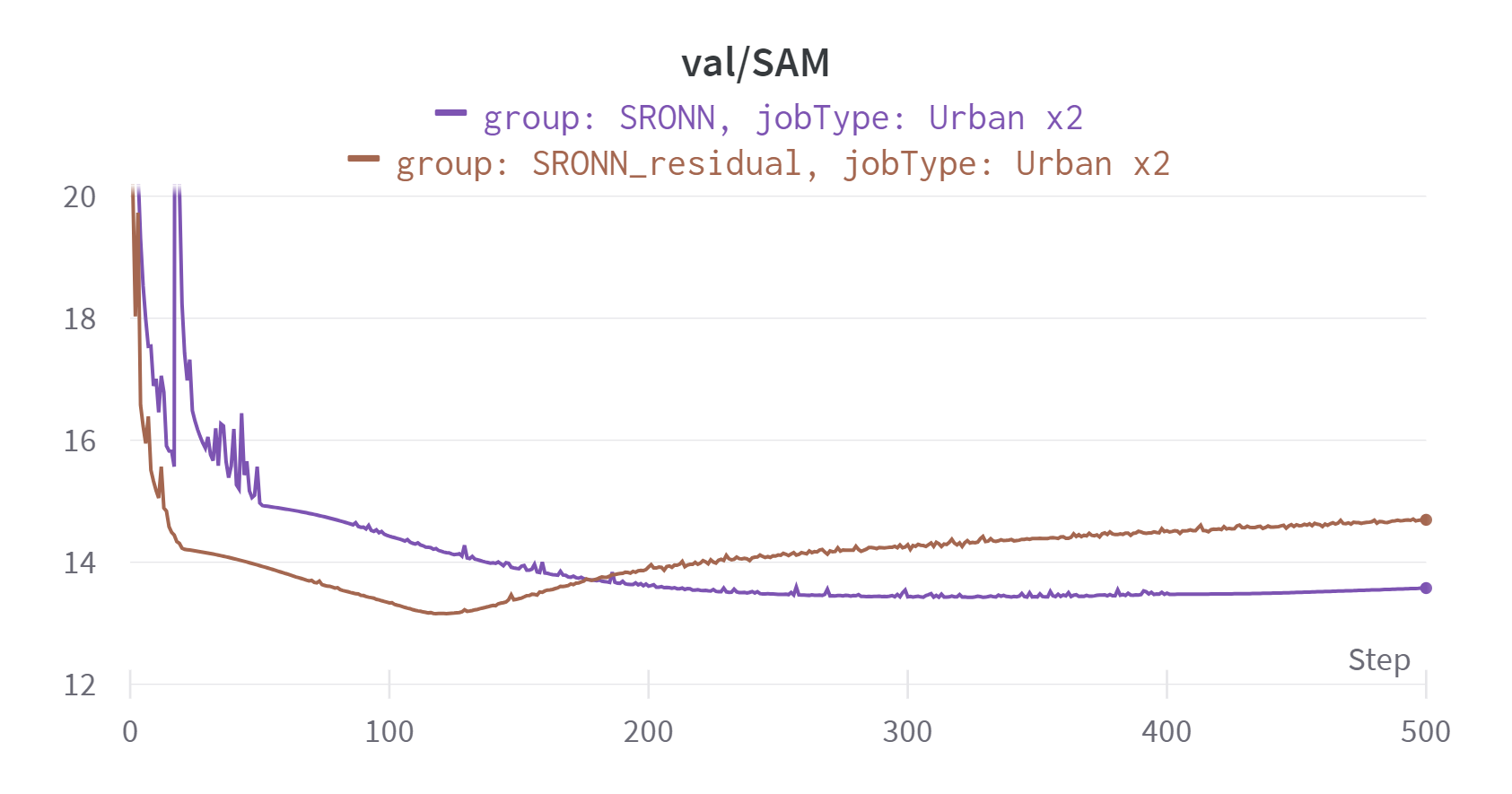}
        \label{fig:Urban_sam}
    }
    
    \caption{SRONN training and validation plots on the Urban dataset with and without a residual connection.}
    \label{fig:Urban_plots}
\end{figure}

\section*{Normalization Results}\label{appendix:norm}

\begin{table*}[h]
    \centering
    \begin{threeparttable}
    	\caption{Normalization Results on Cuprite Dataset.}
   	 \label{tab:cupriteNormResults}
    
    	\begin{tabular}{|l|l|l|l|l|l|l|l|l|}
    	\hline
       	    Model & Residual & Normalization & \# parameters & lr & lr steps & PSNR & SSIM & SAM \\ \hline\hline
            \multirow{2}{4em}{SRCNN} & no & none & 2754976 & $10^{-4}$ & 100k & \textit{27.799} & \textit{0.9766} & 10.136 \\ \cline{2-9}
            ~ & yes & none & 2754976 & $10^{-4}$ & 5k, 40k & 27.783 & 0.9731 & \textit{10.118} \\ \hline\hline
            \multirow{10}{4em}{SRONN} & \multirow{5}{4em}{no} & batch & 8264480 & $10^{-4}$ & 50k & 26.998 & 0.9522 & 10.959 \\ \cline{3-9}
            ~ & ~ & instance & 8264096 & $10^{-4}$ & 25k & 26.248 & 0.9296 & 11.744 \\ \cline{3-9}
            ~ & ~ & l1 & 8264096 & $10^{-4}$ & 50k & 27.506 & 0.971 & 10.438 \\ \cline{3-9}
            ~ & ~ & l2 & 8264096 & $10^{-4}$ & 10k & 27.921 & \textbf{0.9807} & 9.99 \\ \cline{3-9}
            ~ & ~ & none & 8264096 & $10^{-4}$ & 2.5k & 27.882 & 0.9743 & 10.044 \\ \cline{2-9}
            ~ & \multirow{5}{4em}{yes} & batch & 8264480 & $10^{-4}$ & 5k, 40k & 26.968 & 0.9501 & 11.041 \\ \cline{3-9}
            ~ & ~ & instance & 8264096 & $10^{-4}$ & 50k & 27.324 & 0.9626 & 10.662 \\ \cline{3-9}
            ~ & ~ & l1 & 8264096 & $10^{-4}$ & 50k & 27.911 & 0.9761 & 10.005 \\ \cline{3-9}
            ~ & ~ & l2 & 8264096 & $10^{-4}$ & 5k, 40k & \textit{27.939} & 0.9774 & \textit{9.98} \\ \cline{3-9}
            ~ & ~ & none & 8264096 & $10^{-4}$ & 2.5k & 27.927 & 0.9774 & 9.993 \\ \hline\hline
            \multirow{10}{4em}{sSRONN} & \multirow{5}{4em}{no} & batch & 2024816 & $10^{-4}$ & 50k & 27.562 & 0.9689 & 10.371 \\ \cline{3-9}
            ~ & ~ & instance & \textbf{2024720} & $10^{-4}$ & 50k & 26.56 & 0.9501 & 11.286 \\ \cline{3-9}
            ~ & ~ & l1 & \textbf{2024720} & $10^{-4}$ & 50k & 26.448 & 0.9607 & 11.787 \\ \cline{3-9}
            ~ & ~ & l2 & \textbf{2024720} & $10^{-4}$ & 50k & 27.886 & 0.9758 & 10.029 \\ \cline{3-9}
            ~ & ~ & none & \textbf{2024720} & $10^{-4}$ & 15k & 27.863 & 0.9746 & 10.061 \\ \cline{2-9}
            ~ & \multirow{5}{4em}{yes} & batch & 2024816 & $10^{-4}$ & 50k & 27.823 & 0.9732 & 10.104 \\ \cline{3-9}
            ~ & ~ & instance & \textbf{2024720} & $10^{-4}$ & 50k & 27.699 & 0.9701 & 10.242 \\ \cline{3-9}
            ~ & ~ & l1 & \textbf{2024720} & $10^{-4}$ & 5k, 40k & 27.372 & 0.9708 & 10.628 \\ \cline{3-9}
            ~ & ~ & l2 & \textbf{2024720} & $10^{-4}$ & 2.5k, 35k & 27.956 & \textit{0.9775} & \textbf{9.96} \\ \cline{3-9}
            ~ & ~ & none & \textbf{2024720} & $10^{-4}$ & 2.5k & \textbf{27.959} & \textit{0.9775} & 9.961 \\ \hline
    	\end{tabular}
    	\begin{tablenotes}
      	    \footnotesize
      	    \item Bold values are the overall best value for the given metric. Values in italics are the best values for the given model in the absence of a bold value.
        \end{tablenotes}
    \end{threeparttable}
\end{table*}

\begin{table*}[h]
    \centering
    \begin{threeparttable}
    	\caption{Normalization Results on Pavia University Dataset.}
    	\label{tab:paviaNormResults}
    	\begin{tabular}{|l|l|l|l|l|l|l|l|l|}
    	\hline
            Model & Residual & Normalization & \# parameters & lr & lr steps & PSNR & SSIM & SAM \\ \hline\hline
            \multirow{2}{4em}{SRCNN} & no & none & 1306727 & $10^{-4}$ & 5k, 40k & 35.396 & 0.977 & 4.346 \\ \cline{3-9}
            ~ & yes & none & 1306727 & $10^{-4}$ & 2.5k, 10k, 30k & 35.597 & 0.9768 & 4.388 \\ \hline\hline
            \multirow{10}{4em}{SRONN} & \multirow{5}{4em}{no} & batch & 3919975 & $10^{-4}$ & 10k & 34.103 & 0.965 & 6.013 \\ \cline{3-9}
            ~ & ~ & instance & 3919591 & $10^{-4}$ & 20k & 27.385 & 0.8828 & 11.12 \\ \cline{3-9}
            ~ & ~ & l1 & 3919591 & $10^{-4}$ & 50k & 34.475 & 0.9713 & 4.956 \\ \cline{3-9}
            ~ & ~ & l2 & 3919591 & $10^{-4}$ & 5k & 35.16 & 0.9756 & 4.495 \\ \cline{3-9}
            ~ & ~ & none & 3919591 & $10^{-4}$ & 2.5k & 35.857 & 0.9775 & 4.209 \\ \cline{2-9}
            ~ & \multirow{5}{4em}{yes} & batch & 3919975 & $10^{-4}$ & 10k & 34.705 & 0.9688 & 5.242 \\ \cline{3-9}
            ~ & ~ & instance & 3919591 & $10^{-4}$ & 50k & 32.456 & 0.95 & 6.277 \\ \cline{3-9}
            ~ & ~ & l1 & 3919591 & $10^{-4}$ & 50k & 35.828 & 0.9775 & 4.288 \\ \cline{3-9}
            ~ & ~ & l2 & 3919591 & $10^{-4}$ & 10k, 20k, 30k & \textbf{36.069} & \textbf{0.9785} & \textit{4.055} \\ \cline{3-9}
            ~ & ~ & none & 3919591 & $10^{-4}$ & 5k, 40k & 35.914 & 0.9783 & 4.056 \\ \hline\hline
            \multirow{10}{4em}{sSRONN} & \multirow{5}{4em}{no} & batch & 938599 & $10^{-4}$ & 20k & 34.441 & 0.9681 & 5.323 \\ \cline{3-9}
            ~ & ~ & instance & \textbf{938503} & $10^{-4}$ & 50k & 27.792 & 0.8957 & 11.675 \\ \cline{3-9}
            ~ & ~ & l1 & \textbf{938503} & $10^{-3}$ & 50k & 33.884 & 0.9655 & 5.453 \\ \cline{3-9}
            ~ & ~ & l2 & \textbf{938503} & $10^{-4}$ & 50k & 34.934 & 0.9741 & 4.708 \\ \cline{3-9}
            ~ & ~ & none & \textbf{938503} & $10^{-4}$ & 50k & 35.693 & 0.9768 & 4.606 \\ \cline{2-9}
            ~ & \multirow{5}{4em}{yes} & batch & 938599 & $10^{-4}$ & 20k & 35.126 & 0.972 & 4.878 \\ \cline{3-9}
            ~ & ~ & instance & \textbf{938503} & $10^{-4}$ & 50k & 32.559 & 0.9518 & 6.524 \\ \cline{3-9}
            ~ & ~ & l1 & \textbf{938503} & $10^{-4}$ & 50k & 35.672 & 0.9756 & 4.338 \\ \cline{3-9}
            ~ & ~ & l2 & \textbf{938503} & $10^{-4}$ & 10k, 30k & \textit{36.001} & 0.9779 & 4.118 \\ \cline{3-9}
            ~ & ~ & none & \textbf{938503} & $10^{-4}$ & 5k, 40k & 35.926 & \textit{0.9782} & \textbf{4.033} \\ \hline
    	\end{tabular}
    	\begin{tablenotes}
	    \footnotesize
      	    \item Bold values are the overall best value for the given metric. Values in italics are the best values for the given model in the absence of a bold value.
        \end{tablenotes}
    \end{threeparttable}
\end{table*}

\begin{table*}[h]
    \centering
    \begin{threeparttable}
        \caption{Normalization Results on Salinas Dataset.}
        \label{tab:salinasNormResults}
    	\begin{tabular}{|l|l|l|l|l|l|l|l|l|}
    	\hline
            Model & Residual & Normalization & \# parameters & lr & lr steps & PSNR & SSIM & SAM \\ \hline\hline
            \multirow{2}{4em}{SRCNN} & no & none & 2515596 & $10^{-4}$ & 5k & \textit{44.074} & \textit{0.9943} & \textit{1.462} \\ \cline{3-9}
            ~ & yes & none & 2515596 & $10^{-4}$ & 5k, 40k & 44.025 & 0.9941 & 1.517 \\ \hline\hline
            \multirow{10}{4em}{SRONN} & \multirow{5}{4em}{no} & batch & 7546380 & $10^{-4}$ & 5k, 40k & 37.767 & 0.9754 & 3.79 \\ \cline{3-9}
            ~ & ~ & instance & 7545996 & $10^{-4}$ & 50k & 23.106 & 0.7458 & 18.164 \\ \cline{3-9}
            ~ & ~ & l1 & 7545996 & $10^{-4}$ & 50k & 36.285 & 0.9887 & 2.523 \\ \cline{3-9}
            ~ & ~ & l2 & 7545996 & $10^{-4}$ & 3.5k & 38.077 & 0.9918 & 2.082 \\ \cline{3-9}
            ~ & ~ & none & 7545996 & $10^{-4}$ & 2.5k & 43.941 & 0.994 & 1.549 \\ \cline{2-9}
            ~ & \multirow{5}{4em}{yes} & batch & 7546380 & $10^{-3}$ & 50k & 42.656 & 0.9918 & 1.632 \\ \cline{3-9}
            ~ & ~ & instance & 7545996 & $10^{-3}$ & 50k & 32.233 & 0.9529 & 7.342 \\ \cline{3-9}
            ~ & ~ & l1 & 7545996 & $10^{-4}$ & 50k & 43.923 & 0.9937 & 1.422 \\ \cline{3-9}
            ~ & ~ & l2 & 7545996 & $10^{-4}$ & 5k, 40k & 44.12 & 0.9943 & \textbf{1.4} \\ \cline{3-9}
            ~ & ~ & none & 7545996 & $10^{-4}$ & 10k & \textit{44.223} & \textit{0.9944} & 1.461 \\ \hline\hline
            \multirow{10}{4em}{sSRONN} & \multirow{5}{4em}{no} & batch & 1845276 & $10^{-4}$ & 20k & 41.029 & 0.9879 & 2.455 \\ \cline{3-9}
            ~ & ~ & instance & \textbf{1845180} & $10^{-4}$ & 30k & 26.377 & 0.8708 & 11.622 \\ \cline{3-9}
            ~ & ~ & l1 & \textbf{1845180} & $10^{-3}$ & 50k & 34.107 & 0.9801 & 3.147 \\ \cline{3-9}
            ~ & ~ & l2 & \textbf{1845180} & $10^{-4}$ & 50k & 37.75 & 0.9913 & 2.287 \\ \cline{3-9}
            ~ & ~ & none & \textbf{1845180} & $10^{-4}$ & 5k, 40k & 43.558 & 0.9937 & 1.622 \\ \cline{2-9}
            ~ & \multirow{5}{4em}{yes} & batch & 1845276 & $10^{-4}$ & 50k & 42.429 & 0.991 & 1.918 \\ \cline{3-9}
            ~ & ~ & instance & \textbf{1845180} & $10^{-3}$ & 20k & 39.24 & 0.9843 & 2.532 \\ \cline{3-9}
            ~ & ~ & l1 & \textbf{1845180} & $10^{-4}$ & 50k & 43.73 & 0.9935 & 1.446 \\ \cline{3-9}
            ~ & ~ & l2 & \textbf{1845180} & $10^{-4}$ & 5k & 44.039 & 0.9943 & 1.42 \\ \cline{3-9}
            ~ & ~ & none & \textbf{1845180} & $10^{-4}$ & 4.5k, 30k & \textbf{44.286} & \textbf{0.9945} & \textit{1.412} \\ \hline
    	\end{tabular}
    	\begin{tablenotes}
	    \footnotesize
      	    \item Bold values are the overall best value for the given metric. Values in italics are the best values for the given model in the absence of a bold value.
        \end{tablenotes}
    \end{threeparttable}        
\end{table*}

\begin{table*}[h]
    \centering
    \begin{threeparttable}
        \caption{Normalization Results on Urban Dataset.}
        \label{tab:urbanNormResults}
    	\begin{tabular}{|l|l|l|l|l|l|l|l|l|}
    	\hline
            Model & Residual & Normalization & \# parameters & lr & lr steps & PSNR & SSIM & SAM \\ \hline\hline
            \multirow{2}{4em}{SRCNN} & no & none & 2587410 & $10^{-4}$ & 5k, 40k & 25.231 & 0.8878 & 14.811 \\ \cline{3-9}
            ~ & yes & none & 2587410 & $10^{-5}$ & 5k, 40k & \textit{25.872} & \textit{0.8916} & \textit{13.958} \\ \hline\hline
            \multirow{10}{4em}{SRONN} & \multirow{5}{4em}{no} & batch & 7761810 & $10^{-4}$ & 5k, 40k & 22.853 & 0.7566 & 19.631 \\ \cline{3-9}
            ~ & ~ & instance & 7761426 & $10^{-4}$ & 5k, 40k & 20.775 & 0.6761 & 22.462 \\ \cline{3-9}
            ~ & ~ & l1 & 7761426 & $10^{-4}$ & 50k & 25.082 & 0.8675 & 15.33 \\ \cline{3-9}
            ~ & ~ & l2 & 7761426 & $10^{-4}$ & 50k & 25.48 & 0.8905 & 14.332 \\ \cline{3-9}
            ~ & ~ & none & 7761426 & $10^{-4}$ & 5k, 40k & 25.941 & 0.8935 & 13.94 \\ \cline{2-9}
            ~ & \multirow{5}{4em}{yes} & batch & 7761810 & $10^{-6}$ & 50k & 24.558 & 0.8345 & 15.995 \\ \cline{3-9}
            ~ & ~ & instance & 7761426 & $10^{-4}$ & 50k & 23.953 & 0.8437 & 16.478 \\ \cline{3-9}
            ~ & ~ & l1 & 7761426 & $10^{-4}$ & 50k & 26.09 & 0.8959 & 13.515 \\ \cline{3-9}
            ~ & ~ & l2 & 7761426 & $10^{-4}$ & 5k, 40k & \textbf{26.116} & \textbf{0.9023} & \textbf{13.42} \\ \cline{3-9}
            ~ & ~ & none & 7761426 & $10^{-4}$ & 2k & 25.892 & 0.8999 & 13.613 \\ \hline\hline
            \multirow{10}{4em}{sSRONN} & \multirow{5}{4em}{no} & batch & 1899138 & $10^{-4}$ & 5k, 40k & 23.38 & 0.8022 & 18.544 \\ \cline{3-9}
            ~ & ~ & instance & \textbf{1899042} & $10^{-5}$ & 5k, 40k & 19.992 & 0.6549 & 22.723 \\ \cline{3-9}
            ~ & ~ & l1 & \textbf{1899042} & $10^{-4}$ & 50k & 24.352 & 0.8338 & 16.279 \\ \cline{3-9}
            ~ & ~ & l2 & \textbf{1899042} & $10^{-5}$ & 50k & 25.4 & 0.8809 & 14.812 \\ \cline{3-9}
            ~ & ~ & none & \textbf{1899042} & $10^{-4}$ & 3k & 25.818 & 0.8912 & 14.22 \\ \cline{2-9}
            ~ & \multirow{5}{4em}{yes} & batch & 1899138 & $10^{-4}$ & 5k, 40k & 24.53 & 0.8345 & 16.218 \\ \cline{3-9}
            ~ & ~ & instance & \textbf{1899042} & $10^{-4}$ & 5k, 40k & 24.75 & 0.8434 & 15.372 \\ \cline{3-9}
            ~ & ~ & l1 & \textbf{1899042} & $10^{-4}$ & 50k & 25.918 & 0.8895 & 13.783 \\ \cline{3-9}
            ~ & ~ & l2 & \textbf{1899042} & $10^{-5}$ & 50k & 26.019 & \textit{0.8964} & 13.752 \\ \cline{3-9}
            ~ & ~ & none & \textbf{1899042} & $10^{-4}$ & 4k & \textit{26.065} & 0.8963 & \textit{13.681} \\ \hline
    	\end{tabular}
	\begin{tablenotes}
      	    \footnotesize
      	    \item Bold values are the overall best value for the given metric. Values in italics are the best values for the given model in the absence of a bold value.
        \end{tablenotes}
    \end{threeparttable}
\end{table*}

\end{document}